\input epsf

\documentclass[12pt,preprint]{aastex}

\begin{document}
\title{Planetesimal growth in turbulent discs before the onset of gravitational instability}
\author{A. Hubbard\altaffilmark{1,2}, E.G. Blackman\altaffilmark{1,2}}
\affil{1. Dept. of Physics and Astronomy, Univ. of Rochester,
    Rochester, NY 14627; 2. Laboratory for Laser Energetics Univ. of Rochester, Rochester NY, 14623}
\begin{abstract}
It is difficult to imagine a planet formation model that does not at some stage include a gravitationally unstable disc. 
Initially unstable gas-dust discs may form planets directly,  
but the high surface density required has motivated 
the  alternative that gravitational instability occurs 
in a dust sub-layer only after grains have grown large enough
by electrostatic sticking.  Although such growth
up to the instability stage is efficient
for laminar discs,  concern has mounted as to whether realistic 
disc turbulence catastrophically increases the settling time, thereby
requiring additional processes to facilitate planet formation on the needed time scales.
To evaluate this concern, we develop a model for grain growth 
that accounts for the influence of  turbulence
on the collisional velocity of grains and on the scale height of
the dust layer. The relative effect on these quantities depends 
on the grain size. The model produces a disc-radius dependent 
time scale to reach a gravitationally unstable  phase of planet formation.
For a range of dust sticking and disc parameters, 
we find that for viscosity parameters $\alpha \le 10^{-3}$, this time scale is short enough over  a significant range in radii $R$ 
that turbulence does not catastrophically slow
the early phases of planet formation, even in the absence of 
agglomeration enhancement agents  like vortices.

\end{abstract}

\section{Introduction}

Planets are believed to form in the gas and dust discs that  surround  newly formed stars. In principle, 
planets can form purely from gravity, via direct gravitational instability 
in the initial circumstellar disc \citep{Boss97}. However, this mechanism of planet formation 
requires higher initial densities than commonly presumed. 
For a standard MMSN (minimum mass solar nebula) model for the gas and dust discs around a young star, both the initial dust and gas discs are gravitationally stable and the only force available for the early proto-planetesimal
stage of planet formation is electrostatic sticking. 
For planets to form, 
the dust must therefore grow  from electrostatic forces until gravity 
can play a significant role. 
 
The basic model for how such a state could can arise was put forth by Goldreich and Ward (\citet{GW73}, hereafter GW).  In their proposed route to planet formation, dust grains initially collide and stick. The growing grain mass eventually reduces the dust's 
thermal velocity dispersion and therefore, the 
scale height of the dust disc. Eventually the grains  settle into the disc's mid-plane  until the critical density at which this dust disc becomes 
gravitationally unstable is reached and 
the formation of kilometer sized planetesimals is facilitated.  
Subsequent accretion of surrounding gas would then 
complete the core-accretion model for giant planet formation.
Recent observations of CoKu Tau/4 \citep{d'Alessio05} seem to  imply that massive planets
must be able to form within $\sim 10^6$ yr, which is the tightest constraint to date 
on the total available time for any planet formation mechanism.

In the absence of turbulence in the gas disc, the GW model 
is very efficient and the time scale to grow grains to
the size at which enough settling occurs for 
gravitational instability to occur is a small
fraction of $10^6$ years.
 However,  Weidenschilling \citep{Weidenschilling80} argued that turbulence can catastrophically
prevent the required early growth phase 
by stirring up the dust disc, 
thereby delaying or  preventing the subsequent gravitational instability. 
Because of  prevalent
sources of disc turbulence (such as dust-gas shear, \citet{Cuzzi93}, \citet{Champney95} and MRI instability \citet{Balbus91}) a turbulent disc is likely the norm rather than the exception and the GW theory must be revisited.
Ironically, even turbulence driven from the dust settling itself  \citep{Ishitsu03},
might prematurely quench the GW process.  The potential show-stopping effect of turbulence has led to
a substantial body of work incorporating additional physics, such
as anti-cyclonic vortices, that can accelerate the agglomeration
of dust grains should the early GW phase fail. 
But even if  extra processes are present and helpful, 
the need for such processes 
has remained unclear. Despite the conceptual concerns
induced by turbulence, its  actual effect 
on planet growth has not been conclusively calculated.  Regardless of the details of the turbulence, there will exist a dust grain size for which the effects of the turbulence on the dust are 
sufficiently weak that the GW process can proceed to instability.  The question therefore is whether the presence of turbulence catastrophically 
slows or even stops grain growth before that size can be reached.
We attempt to answer that question.

The effect of turbulence on dust growth is 
two-fold. On the one hand it can increase the dust's
 collisional velocities (and so the collisional rate) 
which is a positive effect for growth. Some early work on this was done by \citet{VMRJ} (hereafter VMRJ), further developed by \citet{MMV} (hereafter MMV) and \citet{CH}.
On the other hand, it increases the scale height of the dust (\citet{Dubrulle}, who used the results of VMRJ)
which acts both to decrease the density (and so the collisional rate) and to prevent instability.  Understanding the net effect of
turbulence amounts to understanding its relative effect on
these two parameters as a function of grain size and radial location.  Suchs efforts have been made by \citet{kornet} and \citet{Stepinski} using models by \citet{Cuzzi93}.
In this paper we develop a different model than those of VMRJ or \citet{Cuzzi93} for the effects of turbulence on grain growth and use it to calculate the time evolution of grain growth as a function of dust size, disc location and turbulent strength; so as to determine how long
it takes for the grains to grow large enough to settle
in the disc and reach the gravitationally unstable density.  Both our approach to calculating the effects of turbulence and our results differ from the previous work as will be discussed in section 3.3.
We do not discuss the  physics of the gravitationally unstable and post-gravitationally  
unstable regimes.

In section 2 we derive the basic equation  for  grain
growth, describe the disc model and show how turbulence and
drag on grain motion are included.  In terms of the parameters we develop, the grain growth equation
depends on the (dimensional) ratio of dust collisional velocity and the dust scale height.
In section 3 we derive explicit formulae for these key parameters.
In section 4 we show that there are 5 regimes for the combination
of these parameters and discuss how grain growth proceeds as a progression
through these regimes. In section 5 we collect key results for
the final grain sizes and time scales needed to get to the 
scales at which the disc becomes gravitationally unstable
for fully turbulent discs and for discs with dead zones.
In section 6 we discuss the implications of our results
for constraining the strength of turbulence for which rapid planet 
growth can still occur.  In section 7 we examine our key assumptions and we conclude in section 8.

\section{Basic model of dust grain growth in a turbulent disc}

We consider a protoplanetary disc to be a dust disc embedded in a turbulent gas disc.  At any one time, 
we approximate all dust grains as spherical with identical radii $r_d$, constant density $\rho_d$ and hence dust-dust collisional cross-section $\sigma_{dd} = 4 \pi r_d^2$ and mass $m_d=\frac 43 \pi \rho_d r_d^3$. 
Although the dust in a real disc would have a spectrum of 
sizes, the Central Limit theorem implies that, in the absence of significant destruction from grain-grain collisions, the dust spectrum will rapidly approach a sharply peaked Gaussian.  In section 7.6 we estimate the dust spectrum width upper limit for which the dynamics we model dominate.  In the absence of a detailed sticking and destruction model for collisions 
the extra complication involved with using a spectrum does not guarantee
extra precision for the present calculation.  To encapsulate the unknown physics of dust-dust interactions, 
we presume that a collision between two dust grains has a probability $p$ to result in sticking.  If such a collision does not result in sticking the grains are presumed to be unchanged.  Accordingly we can see that, if $N$ is the dust-dust collision rate then $\frac {dm_d}{dt}=pNm_d$ and so using the equation for $m_d(r_d)$,
\begin{equation}
\frac {dr_d}{dt}=\frac p3 N r_d.
\label{1}
\end{equation}
As dust grain size is a key parameter whose change as a function of the distance from the star we investigate, we will require that dust grains do not significantly migrate radially.

\subsection{Equation of grain growth}

A simple ``particle-in-a-box'' collision model implies that
\begin{equation}
N=\frac{\rho}{m_d} \sigma_{dd} v_c=\frac{\Sigma_d}{H_d m_d} \sigma_{dd} v_c= \frac{3 \Sigma_d}{\rho_d r_d} \frac{ v_c}{H_d},
\label{2}
\end{equation}
where $H_d$ is the dust disc scale height, $\rho$ and $\Sigma_d$ are the dust disc volume and surface densities, 
$\Omega$ is the orbital angular velocity and $v_c$ is the dust-dust collisional velocity. 
The dust and gas scale heights need not be equal, the latter being
given by $H_g=c_s \Omega^{-1}$, where $c_s$ is
the gas thermal speed.  We can also define a velocity $v_h=H_d \Omega$ and write Eq. \ref{2} in terms of a non-dimensional ratio $v_c/v_h$.  Our method of handling turbulence lends itself more readily to Eq. \ref{2} however. 
 
If the velocities of the dust grains are uncorrelated, then their collisional velocity $v_c$ is also their random velocity dispersion and so, akin to the formula for the gas scale height $H_g = c_s/\Omega$, we have that dust scale height $H_d = v_c/\Omega$.  If the ratio $v_c/H_d$ in Eq. \ref{2} is independent of grain size $r_d$,  
then $N \sim  r_d^{-1}$.  For $v_c=\Omega H_d$ and $\rho_d= 1 \rm {gcm}^{-3}$ the growth rate from Eq. \ref{1},
$\frac {dr_d}{dt} \simeq 107 \left( \frac {R}{\rm {1AU}} \right)^{-3} \frac {\rm cm}{\rm yr}$ is independent of dust size and the 
formation of meter sized bodies will occur in $\sim 1$ year.  As without turbulence, gravitational instability of the disc will occur for sub-centimeter sized grains (Eq. \ref{R33End}); this process occurs well within the time scales required by existing constraints (e.g. \citet{d'Alessio05}) even considering settling times.

If, however we allow the velocities of the dust grains to be correlated, as would occur for grains in the same eddy, there is no reason to presume that $v_c=\Omega H_d$ or that $v_c$ and $H_d$ are even closely related.  If the dust couples strongly to the gas and the turbulence 
causes significant mass mixing throught the gas disc, then the gas and dust will share a common scale height $H_d=H_g$.  In addition, very small grains will have the same bulk velocity as any turbulent eddy in which they reside, and  will interact with collisional velocities corresponding to the dust thermal speed, i.e. $v_c=v_{th}$.  The gas sound speed $c_s$ is much greater than the thermal speed of even small dust grains $v_{th}$ so the dust growth rate  would be prohibatively low if such conditions held for large ranges of $r_d$.  

We can, however, imagine that as the dust grains grow and couple less strongly with the gas, the turbulence can collide  
dust grains together with collisional speeds $v_c \sim v_{turbulent}$ that are much greater than $v_{th}$. 
Eventually the growing dust grains will  settle out of the gas disc, 
so that $H_d$ will decrease. For very large dust grains, 
with mean free paths (with respect to each other) 
longer than the largest scale eddies, the effects of the eddies will be uncorrelated, causing $v_c$ to equal $\Omega H_d$, drastically increasing the growth rate 
(Eq. \ref{2}) above the value for very small grains and speeding the onset of gravitational instability.

To facilitate deriving the formulae for $v_c$ and $H_d$ in different regimes, we
define $\phi \equiv \rho_d r_d$, which is proportional to a single
dust grain's surface density. Then,
presuming a constant $\rho_d$, we can combine (\ref{1}) and (\ref{2}) to obtain:
\begin{equation}
\frac {d\phi}{dt}=p \Sigma_d \frac {v_c}{H_d},
\label{main}
\end{equation}
the form of the growth equation we will use for the rest of the paper.  The variable $\phi$ 
will later help clarify 
the effects of different dust grain densities on time scales and critical grain sizes.  As $\frac {\Sigma_d }{H_d}$ is the dust  density in the disc, $\Sigma_d  \frac {v_c}{H_d} dt$ is the surface density of the dust sheet that a dust grain travels through in time $dt$.  Hence (\ref{main}) states that the sticking-parameter modified surface density of that sheet can be added to the penetrating dust grain's surface density, much like a ball moving through cling wrap.

The basic question we seek to answer is 
whether the grain growth is sufficiently rapid that 
the gravitationally unstable stage of planet formation can be reached
long before the total time
scale available for planet formation ($\sim 10^6$ yr, \citet{d'Alessio05}).
In our study, we 
 need to consider that very rapid collisional velocities could be too high for sticking to occur and instead result in the destruction of the grains.
We also need to determine how large the grains must become for them to
settle out and allow gravitational instabilities to occur.  While there exists a dust scale above which the effect of turbulence on the dust velocity dispersion becomes 
sufficiently weak for the dust disc to be gravitationally unstable, that size scale will be far greater than the initial size of the dust, as well as that of the critical size in the absence of turbulence.  A last consideration is our requirement that the growth occurs before significant orbital migration.  

To address these issues, we determine  various regimes for $v_c$ and $H_d$ before the onset of gravitational instability. These regimes 
depend on the quantities $r_d$, $c_s$, the turbulent viscosity parameter $\alpha$ and the mean free path of the gas $l_{mfp}$.
  We then use (\ref{main}) and the values for the ratio of $v_c$ and $H_d$ to calculate the peak collisional velocities and the total growth time before
the onset of  gravitational instability.

\subsection{Disc model}

We assume the disc has  an essentially  isotropic turbulent
viscosity 
\begin{equation}
\nu_T \simeq v_{M}\lambda_M \simeq \alpha c_s H_g,
\label{4a}
\end{equation}
where $v_M$ and $\lambda_M$ are the maximum turbulent velocity and length scales in our assumed Kolmogorov spectrum and 
$\alpha$ is the Shakura-Sunyaev viscosity parameter \citep{SS73}.
Using $\lambda_M/v_M \sim 1/\Omega$, $H_g\Omega=c_s$, and (\ref{4a})
it follows that
\begin{equation}
\lambda_M \simeq \alpha^{\frac 12} H_g
\label{ltmax}
\end{equation}
\begin{equation}
v_M \simeq \alpha^{\frac 12} c_s.
\label{vtmax}
\end{equation}
If $\alpha$ is extremely low, the GW process will proceed as in the absence of turbulence. 
It is likely that  at least dust-gas shear 
(e.g. \cite{Ishitsu03}) 
prevents $\alpha$ from dropping
to values where it could otherwise be ignored.
Our model will involve a progression of dust growth
regimes whose order  requires approximately $\alpha > 2 \times 10^{-6}$.  
Our model needs modification for values of $\alpha$ below that lower bound and if the turbulence is adequately weaker yet that the dust settling time becomes significant then our approach fails completely.  Thus although we are conceptually extending the $\alpha=0$ model of GW, our present techniques do not apply in the turbulence-free limit.

We take a minimum mass solar nebula (MMSN) model with the following scalings from \citet{Sano00}:
\begin{equation}
\Sigma_g(R) = \Sigma_{g0}\left(\frac{R}{1 \rm{AU}}\right)^{-\frac 32}
\label{sigr}
\end{equation}
\begin{equation}
T(R)=T_0\left(\frac{R}{1 \rm{AU}}\right)^{-\frac 12}
\end{equation}
\begin{equation}
c_s(R)=\left(\frac{kT}{m_g}\right)^{\frac 12}=9.9 \times 10^4 \left(\frac{R}{1 \rm{AU}}\right)^{-\frac 14}\left(\frac{\mu}{2.34} \right)^{-\frac 12} \frac {\rm {cm}}{\rm {s}}
\label{cs}
\end{equation}
\begin{equation}
H_g(R)=\frac {c_s(R)}{\Omega(R)},
\label{hg}
\end{equation}
where $\Sigma_{g0} \simeq 1.7 \times 10^3 \rm {g} \rm{cm}^{-2}$ and $T_0=280 \rm{K}$.  We presume that $\Sigma_d = \Sigma_g \times 10^{-2}$,  and $\sigma_{gg}=10^{-15} \rm {cm}^2$ where $\sigma_{gg}$ is the neutral gas-gas cross-section.  The average gas molecule mass $m_g=\mu m_H$ where $\mu=2.34$ is the mean molecular weight of the gas.  The above values are presumed to remain constant in time.

\subsection{Incorporating the role of disc turbulence and drag}

As the dust grains grow, the effect of turbulence on their 
collisional velocity and scale height evolves.  
For dust grains smaller than the mean free path of the gas,
that is $r_d<l_{mfp}=\frac {m_g}{\sigma_{gg} \rho_g}$,
the gas-dust interaction is characterized by Epstein drag, 
whereas for $r_d>l_{mfp}$ we have Stokes drag \citep{Drag}.  Throughout this paper we will use $E$ and $S$ as subscripts refering to behaviour with
Epstein and Stokes drag respectively.  
For a dust grain interacting with a turbulent eddy of size 
$\lambda$ and relative speed $v=v(0)$ at some initial time $t=0$,
the time evolution of $v$ subject to Epstein drag is given by
\begin{equation}
\frac{dv}{dt}=-2 \frac {\rho_g \sigma_{gd} c_s}{m_d} v = -\frac {v}{\tau_E},
\label{dvdtE}
\end{equation}
while for Stokes drag 
\begin{equation}
\frac{dv}{dt}=-\frac {6 \pi r_d c_s \nu}{m_d} v = -\frac{2 \pi l_{mfp} \rho_g r_d c_s}{m_d}v=-\frac{v}{\tau_S},
\label{dvdtS}
\end{equation}
where $\rho_g$ is the gas density, $\sigma_{gd}=\pi r_d^2$ is the gas-dust cross-section and $\nu=\frac{\rho_g l_{mfp}}{3}$ chosen so that the drag equations are equal at the boundary between the Epstein and Stokes regimes $r_d=l_{mfp}$.
We can combine the above two equations by writing 
\begin{equation}
\frac{dv}{dt}=-\frac{v}{\tau},
\label{dvdt}
\end{equation}
where $\tau\equiv Max(\tau_E, \tau_S)$
and 
\begin{equation}
\tau_E=\frac {m_d}{2 c_s \rho_g \sigma_{gd}}=\frac {2 \phi}{3 c_s \frac{\Sigma_g}{H_g}}=\frac {2 \phi}{3 \Sigma_g \Omega},
\label{tauE}
\end{equation}
and
\begin{equation}
\tau_S=\frac{2 \phi^2}{3 l_{mfp} \Sigma_g \Omega \rho_d}.
\label{tauS}
\end{equation}
We note that even though the gas is turbulent, the scale of the smallest 
turbulent eddy is generally  
larger than the scale of the dust grains during the growth phases we consider.
This is why we do not use the high Reynolds number form (\cite{Drag}) of the Stokes
drag (which varies as $v^2$ instead of as $v$).  This will be
justified a posteriori later.

Eqs. (\ref{dvdtE}) and (\ref{dvdtS}) have the same form 
whose solution is 
\begin{equation}
v(t)=v(0)e^{-\frac {t}{\tau}}.
\label{v1}
\end{equation}
where again $\tau\equiv Max(\tau_S,\tau_E)$ 
from (\ref{tauE}) or (\ref{tauS}).
If $v(0)=v_{\lambda}$, the turbulent velocity for an eddy of scale $\lambda$, Eq.  
(\ref{v1}) can be integrated to show that the dust grain will not exit the eddy in a time less 
than the eddy turnover, or destruction, time $t_{\lambda}$.  Upon being released from the eddy at its destruction, 
the grain's speed will have changed by
\begin{equation}
-\Delta v=v(0)-v(t_{\lambda})=v(0)\left(1-e^{-\frac{t_{\lambda}}{\tau}} \right) \simeq v_{\lambda} \left(1-e^{-\frac{t_{\lambda}}{\tau}} \right).
\label{v2}
\end{equation}
For a given stopping time $\tau$, $\Delta v$ from the above equation has its maximum for the largest scale eddies and so
\begin{equation}
\Delta v_M(0)=v_M \left(1-e^{-\frac{t_M}{\tau}} \right)=\sqrt{\alpha} c_s \left(1-e^{-\frac{1}{\Omega \tau}} \right),
\label{vran}
\end{equation}
is the dust's random velocity with respect to the Keplerian disc and the latter equality follows from Eqs \ref{ltmax} and \ref{vtmax}

\section{Calculating the collisonal velocities and scale heights $v_c$ and $H_d$}
To facilitate incorporating the calculations of the previous
sections into  this and the following
sections, we have collected key variables used in this
paper into Table 1.

VMRJ developed an elegant model, later improved by MMV, describing the velocities of dust grains in disks with Komolgorov turbulent spectra.  \citet{Dubrulle} used their results to calculate dust scale heights in such discs.  Unfortunately, VMRJ made recourse to an auto-correlation function (ACF) that was shown by MMV to behave poorly for small time differences (where the most interesting physics occurs).  Neither MMV (their equation 5) nor \citet{CH} strongly motivate their different choices of ACF and we show through direct physical argument that some of their results (and those of \citet{Dubrulle}) are unphysical, raising doubt about the validity of their ACF.  The question of what ACF reproduces our results, if any, is beyond the scope of the present paper.

\begin{figure}
\plotone{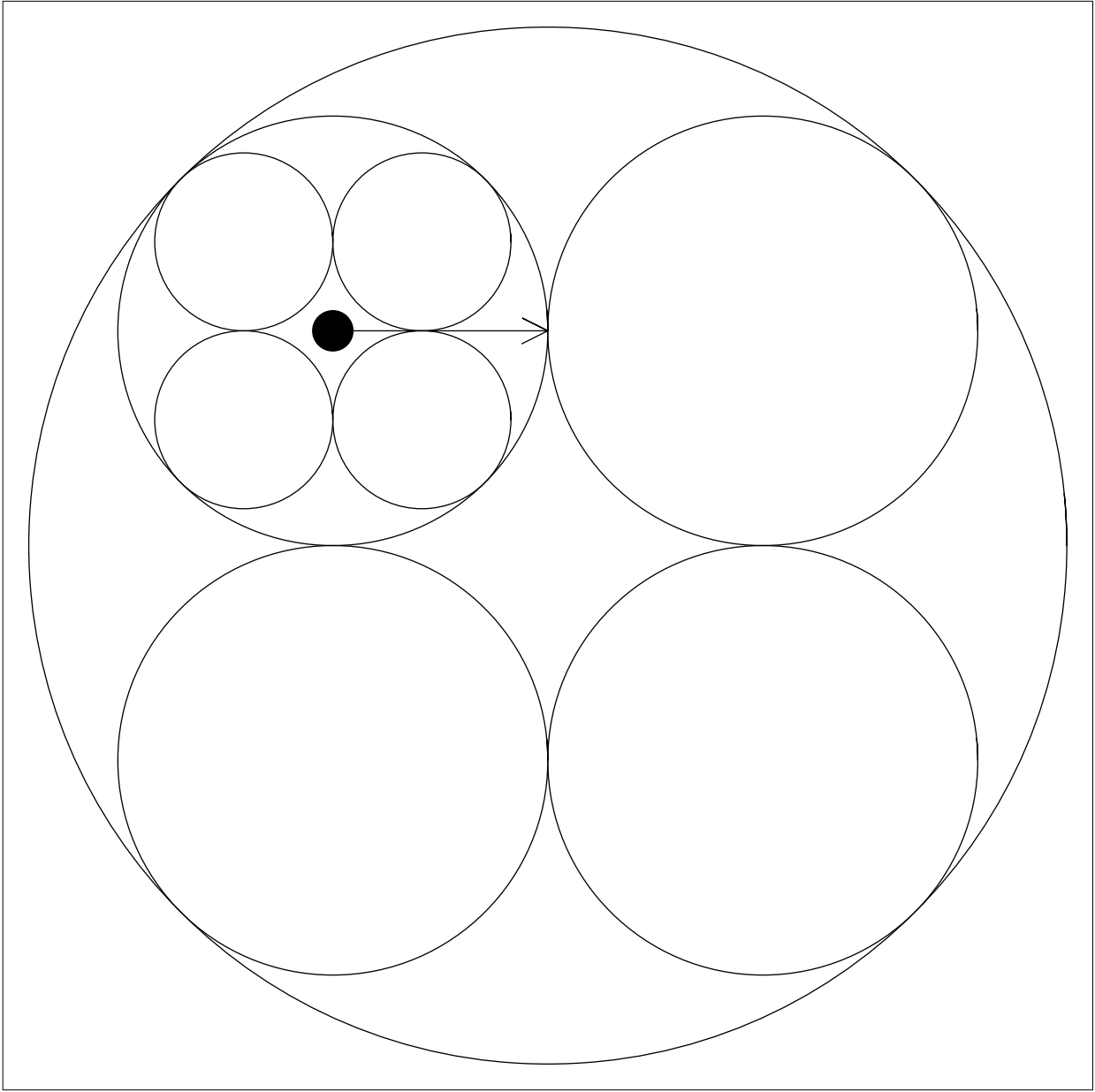}
\caption{A schematic representation of fractally embedded turbulence.  The largest circle is an eddy too large to effect the grain, the smallest circles eddies too weak and the medium circles the eddy scale with the dominant effect.  This is the Goldilocks scale discussed in the text.  As the grain exits one eddy, there is another for it to enter.}
\label{circles}
\end{figure}

  The need for an ACF stems from VMRJ's consideration of the question of a spectrum of dust grains interacting with turbulence.  We instead make the simplifying approximation that all the grains are the same size $r_d$.  For our calculations we note that grains exiting an eddy will promptly enter another eddy of comparable size.
This assumption is straightforwardly justified if the turbulence is characterized by smaller scale eddies  embedded fractally within larger scale eddies.
If instead smaller scale eddies surround larger eddies,  
then justification of the above assumption follows from 
the Kolmogorov hypothesis that eddies of a given size cascade 
by interaction with eddies of comparable size and 
the fact that smaller eddies have lower speeds and less influence on 
a dust grain speed than a larger eddy.
  As shown schematically in Fig. \ref{circles}, there will be large eddies, whose effects on the dust grains are too correlated to contribute significantly to the collisional velocity, small eddies too weak to significantly effect the dust grains and a ``Goldilocks'' scale of eddies ``just right'' to dominate the interaction of dust grains.  By making the simplifying assumption that all the grains have the same size, it is a simple task to identify the Goldilocks eddy scale as a function of grain size and calculate its effects.

\subsection{Collisional Velocity of Grains}

To estimate the characteristic collisional velocity between 
two dust grains we assume that the grains' individual 
speeds are each $\Delta v$ from (\ref{v2}), 
determined by having  exited separate eddies of comparable size.
We then estimate the collisional velocity $v_c$ as the average relative speed
of dust grains 
 from head-on and catch-up collisions:
\begin{equation}
v_c(0)\simeq  
{1\over 2}(\Delta v(0)+\Delta v(0)) + {1\over 2}(\Delta v(0)-\Delta v(0)) 
=\Delta v(0)= v_{\lambda}\left(1-e^{-\frac{t_{\lambda}}{\tau}} \right) 
\label{v0},
\end{equation}
where the latter equality follows 
from (\ref{v2}).  Since $\Delta v$ 
represents  a linear combination of two dust  grain speeds, 
and both sides of  (\ref{dvdtE}) and
(\ref{dvdtS}) above are linear, we then know from (\ref{v1}) and
(\ref{v0}) that $v_c$ evolves
as
\begin{equation}
v_c(t)=v_c(0) e^{-\frac{t}{\tau}} = v_{\lambda}\left(1-e^{-\frac{t_{\lambda}}{\tau}} \right) e^{-\frac{t}{\tau}}.
\label{vt}
\end{equation}
Averaging (\ref{vt}) over an eddy turnover time 
from $t=0$ to $t_{\lambda}$ then gives 
\begin{equation}
v_c(x)=\frac {1}{t_{\lambda}} \int_0^{t_{\lambda}} v_c(t) dt=\frac {v_{\lambda}}{x} \left(1-e^{-x} \right)^2,
\label{vcfirst}
\end{equation}
where $x=\frac{t_{\lambda}}{\tau}$.  For a Komolgorov turbulent spectrum, $v_{\lambda}=v_M \left(\frac{\lambda}{\lambda_M}\right)^{\frac13}$ and $\lambda=v_{\lambda} t_{\lambda}$ which we can use with (\ref{ltmax}) and (\ref{vtmax}) to write:
\begin{equation}
v_{\lambda}=\left(\tau \Omega x \right)^{\frac 12} \sqrt{\alpha} c_s.
\label{vlambda}
\end{equation}
Together, Eqs \ref{vcfirst} and \ref{vlambda} give
\begin{equation}
v_c(x)=\left(\tau \Omega \right)^{\frac 12} \frac {\left(1-e^{-x} \right)^2}{\sqrt{x}} \sqrt{\alpha} c_s .
\label{vcfinal}
\end{equation}
The Goldilocks eddy scale is the one that maximises $v_c(x)$.  Equations \ref{vc2}, \ref{vc1}, \ref{vc0} and \ref{vc3} below describe the dust-dust collisional velocity $v_c$ in four different size regimes and are gathered in Table \ref{RegimeOrder}.

\subsubsection{Maximuma of $v_c(x)$}

From Eq. \ref{vcfinal}, maximizing $v_c(x)$  gives
\begin{equation}
Max[v_c(x)]=v_{c,2}\equiv B \sqrt{\tau \Omega} \sqrt{\alpha} c_s 
\label{vc2}
\end{equation}
with $B=0.53$ and the maximum occurs at $x=x_{max}
=2.3$.  As long as there exists a $\lambda$ for which $x=x_{max}$ we will use Eq. \ref{vc2} as the turbulence induced collisional velocity.  However, a Komolgorov turbulent spectrum has both a maximum and a minimum scale such that $t_M>t_{\lambda}>t_m$ where $t_M=\frac {\sqrt{\alpha}H_g}{\sqrt{\alpha}c_s}= \frac {1}{\Omega}$ is the turnover time for the largest scale eddy and $t_m=\sqrt{\frac {l_{mfp}}{3 \alpha c_s \Omega}}$ is that for the smallest.  We seperately treat those dust sizes for which Goldilocks eddy scales do not exist.

\subsubsection{Small $\tau$ cases}

For small enough $\tau$, the minimal value of $x=\frac {t_m}{\tau}>x_{max}$ and we cannot use Eq. \ref{vc2}.  However for such large $x$, $\left(1-e^{-x} \right)^2$ is close to $1$ and we can approximate (\ref{vcfinal}) as:
\begin{equation}
v_{c}(x)=v_{c,1}(x)\equiv \left(\frac {\tau \Omega}{x} \right)^{\frac 12} 
\sqrt{\alpha} c_s =\left(\frac {\Omega}{t_m} \right)^{\frac 12}\tau \sqrt{\alpha} c_s. 
\label{vc1}
\end{equation}
This approximation is useful 
because the time scale of the regime (regime $1.1$) where we will use it is only weakly dependent on its final size as can be seen in Table \ref{RegimeTime}.

For sufficiently small values of $\tau$, the values of $v_c$ from (\ref{vc1}) will be less even than the dust's thermal speed.  For such grains we then
just use the dust thermal velocity for $v_{c}$, 
\begin{equation}
v_c=v_{c, 0}(\phi)=v_{c, th}(\phi)\equiv c_s\sqrt{\frac{m_g}{m_d}}=
\sqrt{\frac{3 \rho_d^2 m_g}{4 \pi \phi^3}},
\label{vc0}
\end{equation}
where $m_g$ is the molecular weight of the gas.

\subsubsection{Large $\tau$ case}

For large enough $\tau$, the maximal value of $x=\frac {t_M}{\tau}= \frac {1}{\Omega \tau } < x_{max}$ and again we cannot use (\ref{vc2}).  In this regime we have
\begin{equation}
v_c=v_{c, 3}\equiv v_M \frac {1}{x} \left(1-e^{-x} \right)^2=c_s\sqrt{\alpha} \Omega \tau \left(1-e^{-\frac{1}{\Omega \tau}} \right)^2,
\label{vc3}
\end{equation}
as that maximizes Eq. \ref{vcfinal} given the constraints on $x$.

\subsubsection{Maximal $v_c(\tau)$}

Equations (\ref{vc1}), (\ref{vc2}) and (\ref{vc3}) allow us to maximise $v_c$ with respect to $\tau$.  We find
\begin{equation}
Max [v_c(\tau)]= v_{c, \tau}\equiv C \sqrt{\alpha} c_s, 
\label{vcmax}
\end{equation}
for $C=0.407$ at $x=1.26$, which is less than $x_{max}$.  Accordingly, the maximum $v_c(\tau)$
 occurs   in the regime for which
 $v_c=v_{c, 3}$.

\subsection{Scale height $H_d$}

In addition  to affecting the collisional velocities, turbulence  also affects the dust scale height
$H_d$ by preventing settling.  We can use Eq. \ref{dvdt} 
to calculate the settling time of a dust grain in the disc.  
Ignoring turbulence for the moment,  a grain at height $z$ above the midplane moving towards the midplane with speed $v=\frac {dz}{dt}$ feels a vertical force 
\begin{equation}
\frac {d^2z}{dt^2}=-\frac{GM}{R^2}\frac{z}{R}-\frac {1}{\tau}\frac{dz}{dt}
\label{vertforce}
\end{equation}
which gives a terminal velocity ($\frac{d^2z}{dt^2}=0$) of
\begin{equation}
\frac {dz}{dt}=-\frac{GM}{R^3} \tau z=-\Omega^2 \tau z.
\label{dzdt}
\end{equation}
Accordingly, $z(t)=z(0)e^{-\frac{t}{t_s}}$, where the settling time $t_s=\left(\Omega^2 \tau \right)^{-1}$.

In the presence of turbulence however, there will be ``updrafts'' of turbulent eddies supporting the grains against the gravity.  Balancing the force of gravity with the force of a largest scale eddy on a grain at rest gives:
\begin{equation}
\Omega^2 z = \frac {v_M}{\tau} = \frac{\sqrt{\alpha}c_s}{\tau},
\label{forceb1}
\end{equation}
which we rewrite as
\begin{equation}
z= \frac{\sqrt{\alpha}}{\Omega \tau} H_g.
\label{forceb2}
\end{equation}
As long as $\sqrt{\alpha}/(\Omega \tau) < 1$, force-balance will occur for $z<H_g$ and hence we can use $z$ from (\ref{forceb2}) as $H_d$.  For smaller grains however, implicit assumptions of isotropy of turbulence break down at the boundaries of the gas disc and so we limit the scale height of the dust to that of the gas $H_d=H_g$.  Accordingly we have that for $\sqrt{\alpha}/(\Omega \tau)>1$, the dust does not settle and:
\begin{equation}
H_{d, 1}=H_g,
\label{hd1}
\end{equation}
while for $\sqrt{\alpha}/(\Omega \tau)<1$ the dust settles and instead:
\begin{equation}
H_{d, 2}=\frac{\sqrt{\alpha}}{\Omega \tau} H_g=\frac{\sqrt{\alpha}}{\Omega^2 \tau} c_s.
\label{hd2}
\end{equation}
The results are gathered in Table \ref{RegimeOrder}.  It is worth noting that the above is well approximated by the distance the largest scale eddies (which have the greatest effect on the dust's movement with respect to the Keplerian disc) can carry a grain in a single settling time, bounded above by the scale height of the gas disc and below by the distance a single largest scale eddy can carry a grain.

 In our calculations of  $H_{d, 2}$ we have implicitly 
assumed that the grain growth time scale is longer than the settling time scale.  If we calculate the growth time scale $\frac{\phi}{\dot{\phi}}$ and the settling time $t_s$ at the onset of settling ($\tau=\frac{\sqrt{\alpha}}{\Omega}$), for Epstein drag we find that the settling time is shorter by a factor of $\frac{2 p B \alpha^{-\frac14}}{3} \frac{\Sigma_d}{\Sigma_g}$.  For $\alpha>10^{-8}$ then, our assumption is justified.

\subsection{Comparison to previous work}

MMV improved on the work of VMRJ by noting that the cut-off of turbulence at small scales effects the dust's velocities, and found all the regimes we have (though they ignored the possibility of thermal interactions).  However, the behaviour they found in the regime that corresponds to our section 3.1.2 differs from our results.  In this regime, the velocities of dust grains inside a given eddy will be extremely correlated because the grains are very small.  It follows that their interactions will be purely thermal except inside the thin ``skin'' layer of each eddy into which dust grains can penetrate before losing any initial velocity relative to the eddy.  We can approximate this by sending a test grain into such an eddy with initial velocity $v$.  The test grain will interact with those dust grains already trapped by that eddy with its initial velocity for its stopping time $\tau$ before decellerating and becoming trapped.  As the grain will remain trapped for an eddy turnover time $t_{\lambda}$, the grain's average collisional velocity will be:
\begin{equation}\
v_{average} \sim \frac{v\tau+0 \times (t_{\lambda}-\tau)}{t_{\lambda}} \simeq \frac {v \tau}{t_{\lambda}}.
\label{minapprox}
\end{equation}
The above physical estimate not only scales with $\tau$ as (\ref{vc1}) but the two equations are in fact equal.  Because the results of MMV show a different and unphysical scaling in this regime, we find it likely that their ACF and its implications are unphysical.

As the approach of \citet{kornet} and \citet{Stepinski} (using the results of \citet{Cuzzi93}) involves averaging over the turbulent energy spectrum, their results are only weakly dependent on the small scale turbulence and they do not consider this extremely correlated regime just discussed for the collisional velocity.  It is worth noting that \citet{Cuzzi93} uses a non-Kolmogorov turbulent energy spectrum \citep{meek}.

Another difference between our work and that of VMRJ or \citet{kornet} is in the behaviour of large ($\tau \Omega >1$) dust grains, for which settling occurs.  Our calculation of the dust scale height differs from \citet{Dubrulle} (using the work of VMRJ) and \citet{kornet} who find the dust scale height behaving as $\tau^{-\frac 12}$ while ours behaves as $\tau^{-1}$.  As we derived our scale height from direct considerations of force-balance however, we feel justified in using our results.

\section{Regimes of the equation for grain growth}

In the previous sections we have derived a 
progression of four formulae 
for $v_c$ and two for $H_d$ 
that apply as 
 dust grains grow, as gathered in Table \ref{RegimeOrder}.  As long as the dust grows monotonically, 
for any values of $R$ 
and $\alpha$ we have a progression of up to five regimes for the ratio $\frac {v_c}{H_d}$ and hence Eq. \ref{main}.   For $R$ and $\alpha$ in
the allowed range of this paper ( $2 \times 10^{-6}<\alpha<10^{-2}$ and $0.5AU<R<8AU$), Table \ref{RegimeOrder} summarizes
 the regimes which we label by 
 $i.j$ where $i$ corresponds to  $v_{c,i}$ and $j$ 
 corresponds to  $H_{d,j}$.  Table \ref{RegimeOrder} can be used to calculate the grain sizes $\phi$ for which regime transition occurs, and to solve Eq. \ref{main} for the time it takes for the grains to grow through each regime into the next.  As these calculations are straightforward, tedious, and repeated for 5 regimes, they are collected in the Appendix 1.  Tables \ref{RegimeSize} and \ref{RegimeTime} contain the results of these calculations.

As our time scale calculations involve ratios of $v_c$ and $H_d$ we can see that once the grains begin to settle (regimes $2.2$ and $3.2$), the factors of $\alpha$ will cancel in Eq. (\ref{main}), resulting in the lack of $\alpha$ dependence of the time scales seen in Table \ref{RegimeTime}. However, as seen in Table \ref{RegimeSize}, 
the grain size range over which the regimes occur still depends 
on $\alpha$ because the beginning and conditions for each regime depend
on $\alpha$. Thus the time scale for each regime, and the overall growth
time to the gravitational instability phase still depends on $\alpha$.

\section{Results for growth time scales and sizes}

\subsection{Grain sizes at the end of each regime}

\begin{figure}
\vspace{-.1cm} \hbox to \hsize{ \hfill \epsfxsize7.85cm
\epsffile{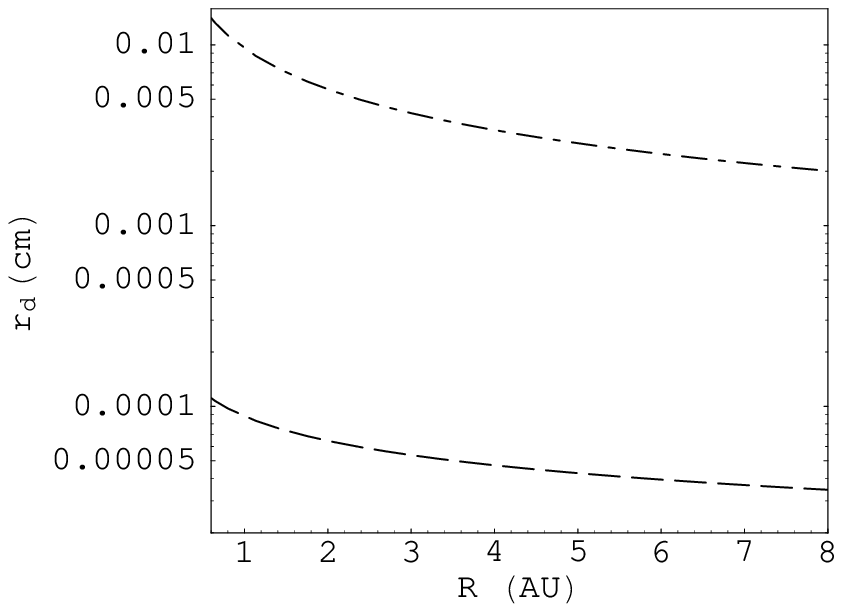} \epsfxsize7.5cm \epsffile{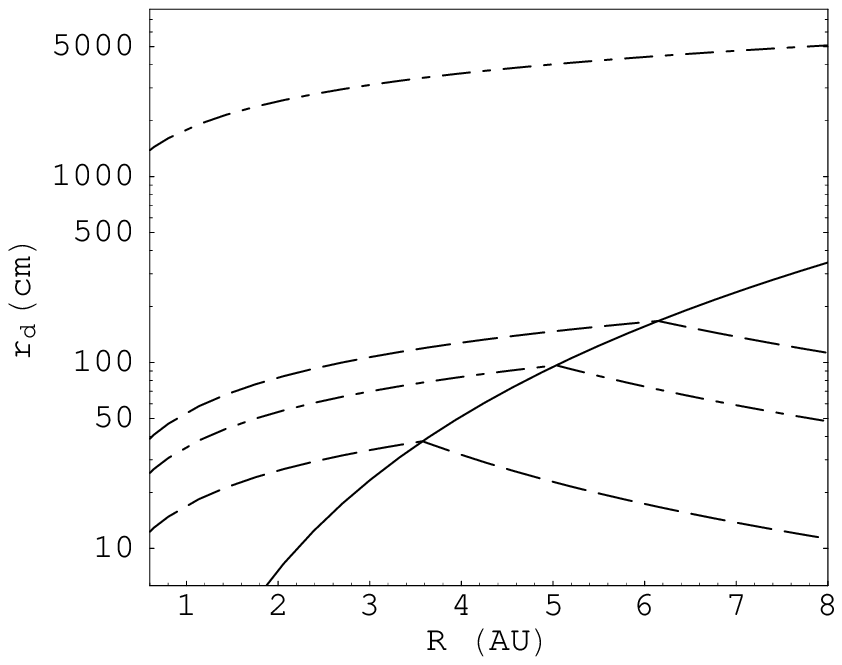} \hfill }
\vspace{-.1cm} \hbox to \hsize{ \hfill \epsfxsize7.8cm
\epsffile{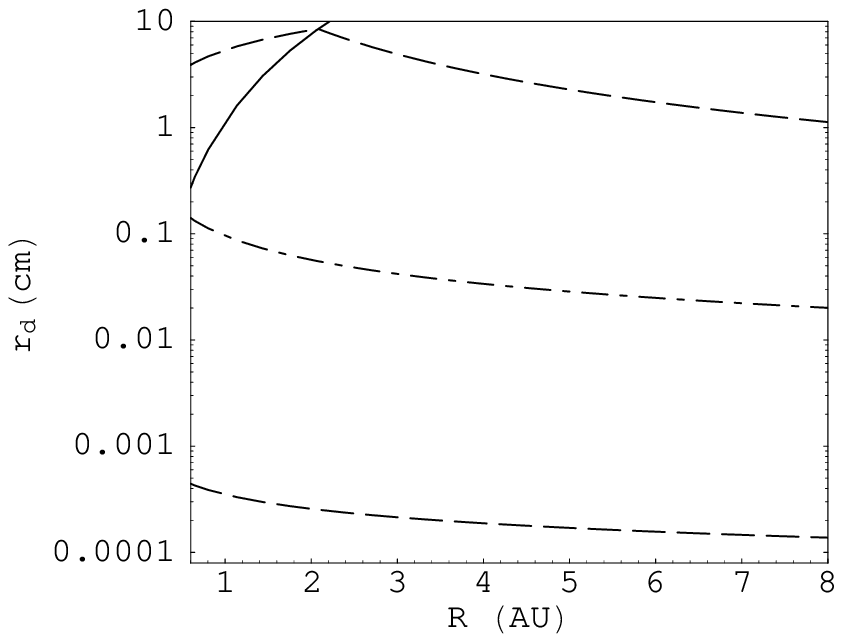} \epsfxsize7.5cm \epsffile{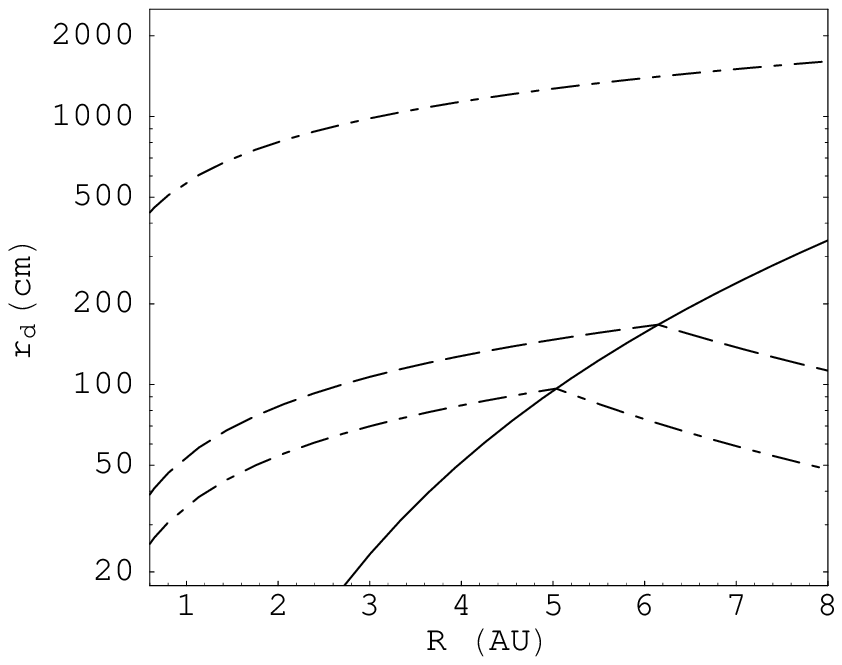} \hfill }
\caption{Size of the dust grains at regime boundaries  
as a function of the distance from the star for $\rho_d=1 \rm{gcm}^{-3}$.  The top pair of panels are for $\alpha=10^{-2}$ while the bottom pair are for $\alpha=10^{-4}$.  The solid curve is the dividing line between Epstein (right) and Stokes (left) drag.  It lies above the plotted region for the top left panel.
The dashed and dot-dashed curves represent the grain sizes for which either the collisional velocity of the grains or the dust disc scale height changes its dependence on the drag stopping time $\tau$ (see Table \ref{RegimeOrder}).  In order of increasing grain size, the dashed and dot-dashed curves denote the regime transitions between regimes $0.1$ and $1.1$, $1.1$ and $2.1$, $2.1$ and $2.2$, $2.2$ and $3.2$, and the curve $\tau=\Omega^{-1}$.  The highest curve is the grain size at which gravitational instability occurs.}
\label{SizeBase}
\end{figure}

Figure \ref{SizeBase} shows the grain sizes calculated from the formulae
for $\phi_f$ for the end of each regime of Table \ref{RegimeSize} 
as a function of disc location.  The calculations use a fixed $\rho_d= 1 \rm {gcm}^{-3}$, so $r_d =r_d(\phi) \propto \phi$ . 
The regimes are labeled in the Figure captions. The solid line is the curve $\rho_d l_{mfp}$, which represents
the boundary between Stokes (left) and Epstein (right) drag regimes.

An important feature of Fig. \ref{SizeBase} is that there is a  maximum grain size in  each regime
which falls along the line bounding the Stokes and Epstein
drag regimes. This arises because all of the equations for $\phi_f$ of the previous sections
that define
the boundaries between regimes are increasing functions of $R$ for 
Stokes drag and decreasing functions of $R$  for Epstein drag.  
Because the drag viscosities are equal at the boundary between Stokes and Epstein
regimes (Sec. 2.3),  for each regime of Sec. 4,
 there is a triple point in the plots of $\phi_f$ vs $R$
 where $\phi_E=\phi_S=\rho_d l_{mfp}$. The triple 
point is formed by the intersection of
the curves corresponding to Epstein and Stokes drags,
and the line which represents this boundary as a function of $R$.  Compared
to the Stokes regime, the Epstein regime curve is less dependent on the various parameters such as $\rho_d$ and $l_{mfp}$.  

The triple point of the curve $\tau=\Omega^{-1}$
is particularly significant because that curve defines the grain size scale for which grain infall from the drag
on the gas is important (discussed in Sec. 7.2).  That this is where the infall is maximized is not surprising as it occurs when all of our time scales, the stopping time $\tau$, the orbital time $\Omega^{-1}$ and the settling time $t_s$ are equal.
For our disc model this triple point is independent of $\alpha$ and 
is given by 
\begin{equation}
R_{tp} = 6.14 \left(\frac{\Sigma_{g0}}{1.7 \times 10^3 \rm{gcm}^{-2}}\right)^{\frac{8}{17}} \left(\frac{\rho_d}{1 \rm{gcm}^{-3}}\right)^{-\frac {4}{17}} AU.
\label{RTriple}
\end{equation}

Although we first present results without including dust migration,
note that due to the monotonic behaviour of the boundaries in a given drag regime, inward migration 
would cause grains of a constant size to rise above the size curves of Figs. 1 and 2 in the Stokes regime but fall below the curves in 
the Epstein regime. As a grain grows from the smallest  to the largest $\phi$ within a regime,
 it will experience Epstein drag for $R$ above the triple point of the regime, Stokes drag if $R$ is less than the triple point of the prior regime, and it will transit from Epstein to Stokes drag
if $R$ lies between the two triple points. 
Figure \ref{SizeBase} 
shows that  because of this behaviour, the peak grain size occurs for $5<R<8 \rm{AU}$ for dust grains of $r_d\sim$  several meters.  This is the size at which gravitational instability takes over and marks the transition to growth regimes subsequent to those studied in this paper.


\subsection{Total time scales to grow to gravitationally unstable regime}

\begin{figure}
\plottwo{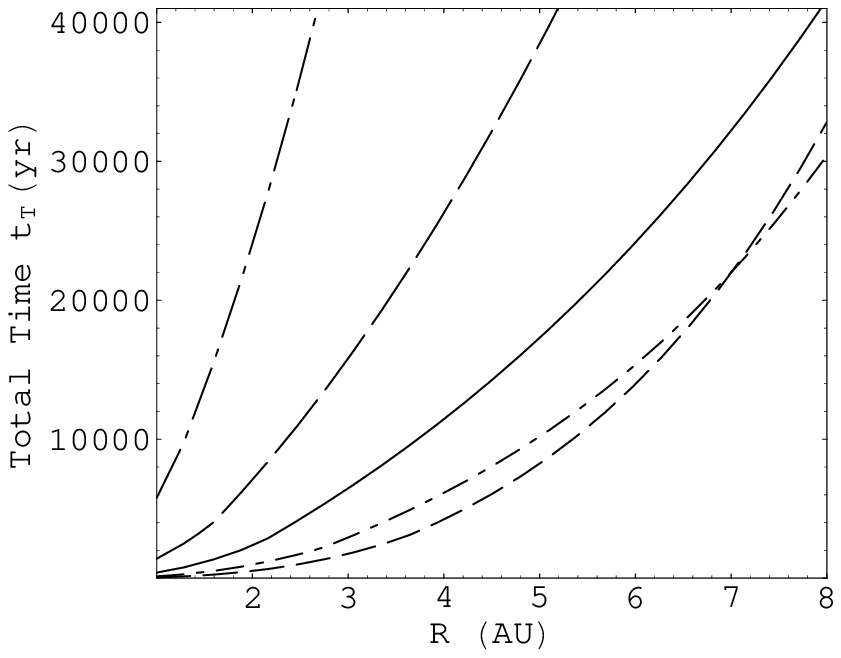}{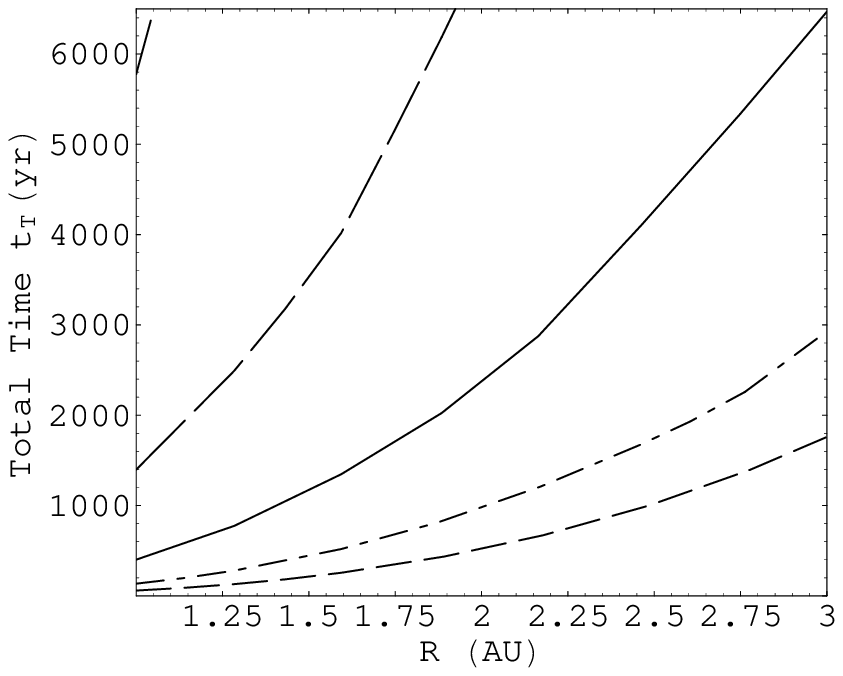}
\caption{Total time scales to grow through all regimes from Table \ref{RegimeOrder} 
up to the gravitionally unstable regime
for various values of $\alpha$  as a function of distance from the star.  
The solid line is $\alpha=10^{-4}$.  The short dashed and dot-dashed lines are $\alpha$ of $10^{-2}$ and $10^{-3}$ respectively while the long dashed and dot-dashed lines are for $\alpha$s of $10^{-5}$ and $10^{-6}$.  For any $R$ there is a value of $\alpha$ that minimizes the total time scale.  
We have used $\rho_d=1 \rm{ gcm}^{-3}$ and $p=1$ for these plots.  The various time scales vary inversely with $p$ ($t \sim 1/p$).}
\label{TimeLargeSmall}
\end{figure}

\begin{figure}
\plottwo{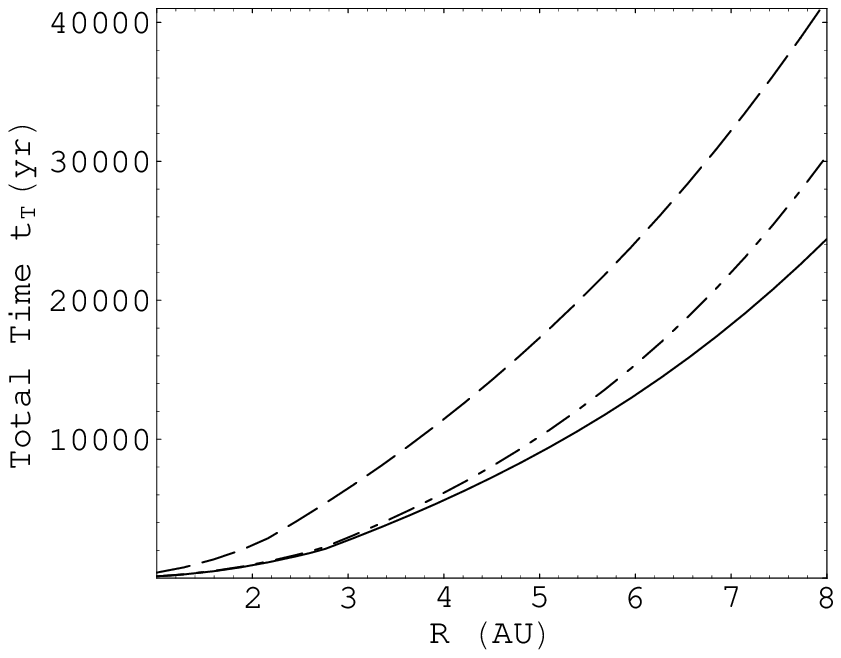}{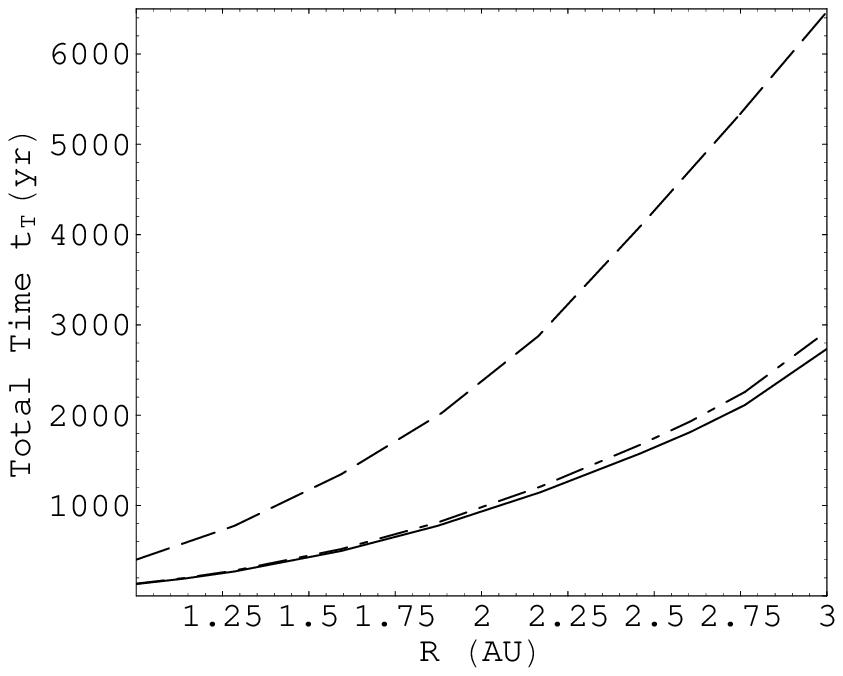}
\caption{The time scale for growth through all the regimes to gravitational instability as in Fig. (\ref{TimeLargeSmall}),
but for a disc with live zone $\alpha=10^{-3}$ and a dead zone with $\alpha=10^{-4}$ (solid line).  Also shown are   the time scales for discs of uniform $\alpha= 10^{-3}$ (dash-dotted) and $\alpha=10^{-4}$ (dashed).    The various time scales vary inversely with $p$ ($t \sim 1/p$).}
\label{TimeLiveDead}
\end{figure}

\begin{figure}
\vspace{-.1cm} \hbox to \hsize{ \hfill \epsfxsize7cm
\epsffile{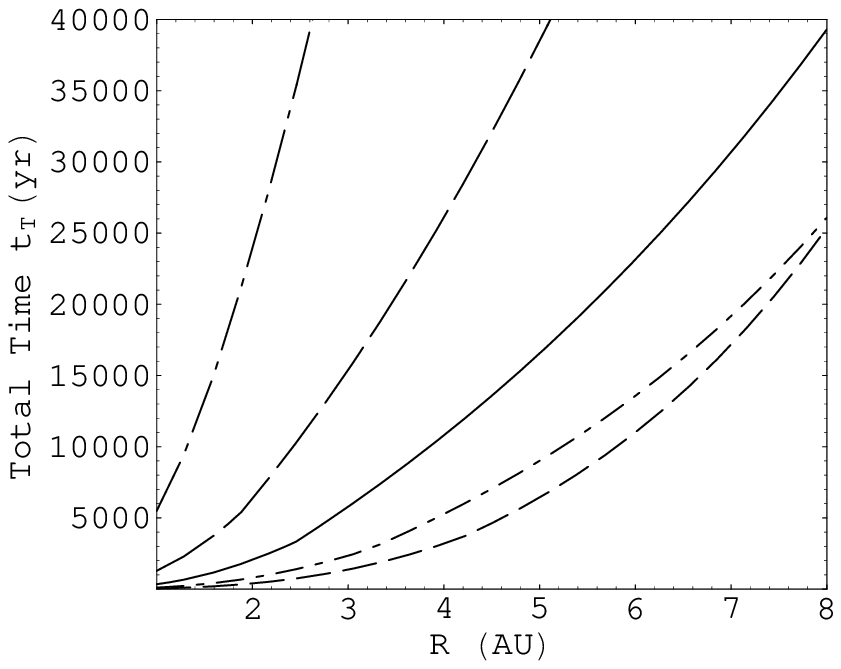} \epsfxsize7cm \epsffile{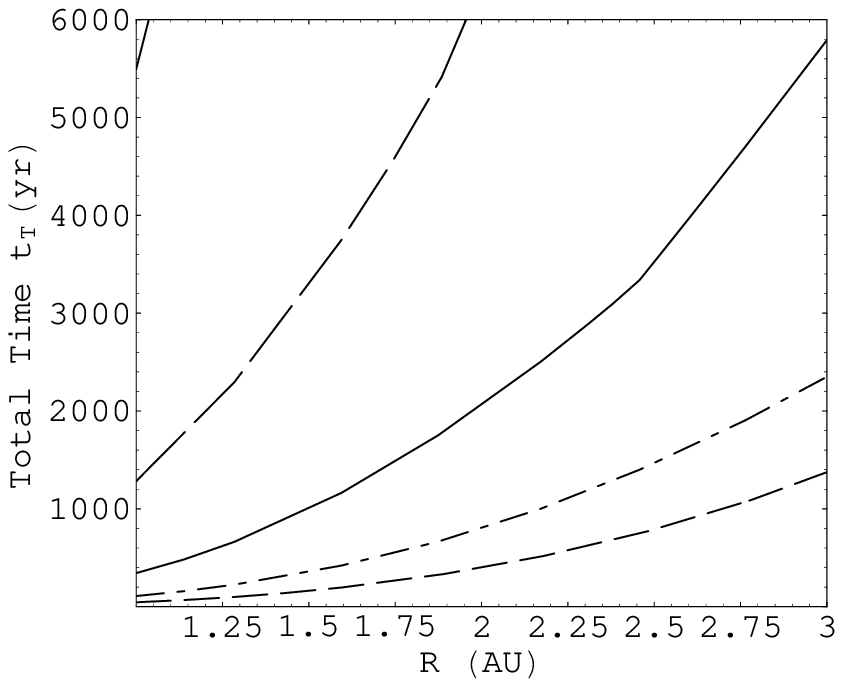} \hfill }
\vspace{-.1cm} \hbox to \hsize{ \hfill \epsfxsize7cm
\epsffile{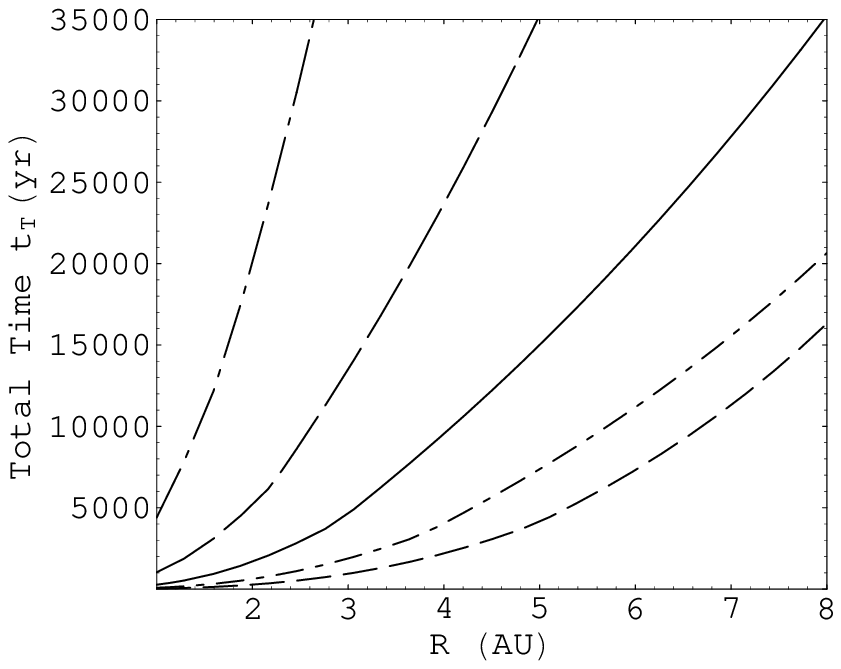} \epsfxsize7cm \epsffile{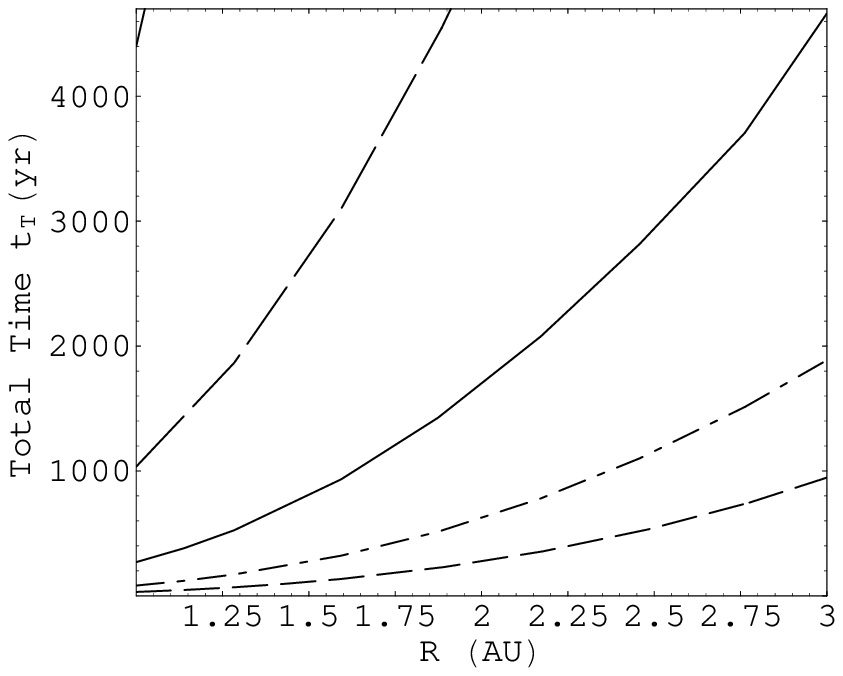} \hfill }
\vspace{-.1cm} \hbox to \hsize{ \hfill \epsfxsize7cm
\epsffile{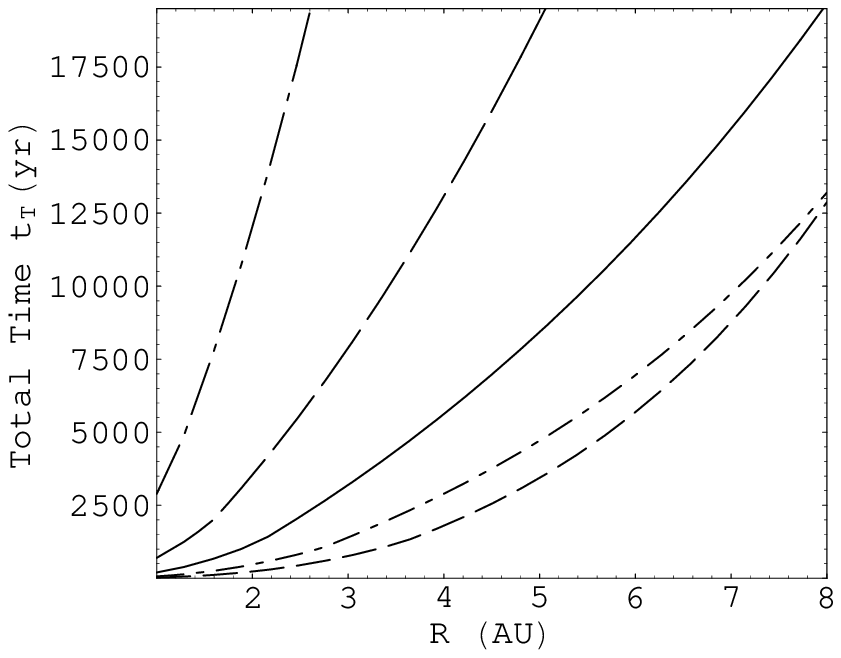} \epsfxsize7cm \epsffile{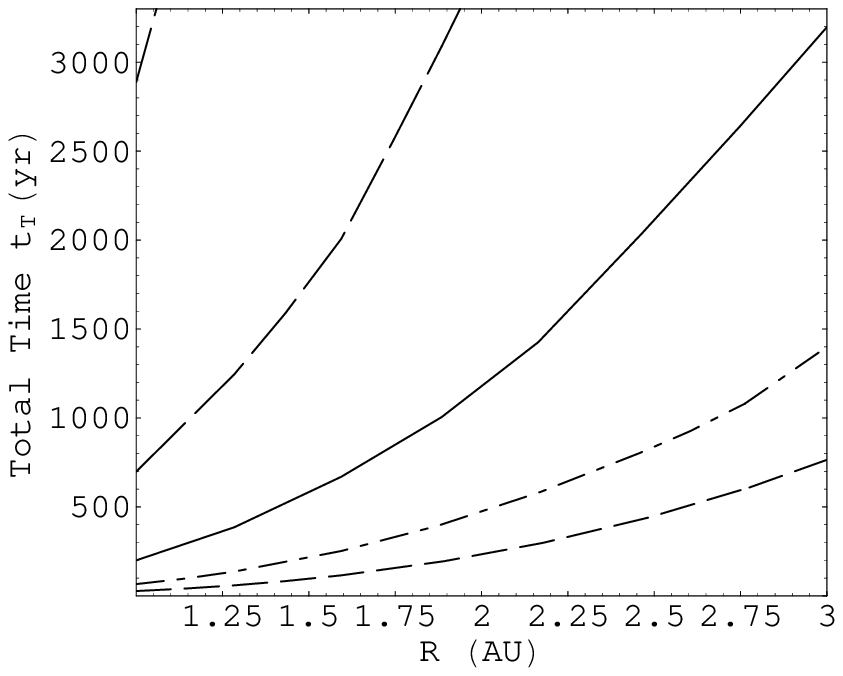} \hfill }
\caption{The time scales for growth through all the regimes to gravitational instability as a function of the distance from the star and for a range of parameters.  We use parameters from our MMSN (section 2.2) and set the dust grain density $\rho_d=1 \rm{gcm}^{-3}$ except as listed below.
The top row halves the dust grain density ($\rho_d=0.5 \rm { gcm}^{-3}$).  The middle row doubles the surface density of both the gas and dust discs ($\Sigma_g(1\rm{AU})=2\Sigma_{g0}$, $\Sigma_d/\Sigma_g=0.01$).  The bottom row doubles the ratio of dust to gas in the disc ($\Sigma_d/\Sigma_g=0.02$).  As in Fig. \ref{TimeLargeSmall} the solid lines are $\alpha=10^{-4}$,  the short dashed and dot-dashed lines are $\alpha$ of $10^{-2}$ and $10^{-3}$ respectively while the long dashed and dot-dashed lines are for $\alpha = 10^{-5}$ and $\alpha=10^{-6}$.  In all panels we have optimal sticking ($p=1$).  The time scales vary inversely with the sticking parameter $p$ ($t \sim 1/p$).}
\label{TimeVarious}
\end{figure}

In figures 
\ref{TimeLargeSmall},
\ref{TimeLiveDead} and
\ref{TimeVarious}
we plot the total time scale $t_T$ to grow through all
regimes 0.1 through 3.2 and thus the total
time to reach the gravitationally unstable phase as calculated from Tables \ref{RegimeSize} and \ref{RegimeTime}.  A key feature to note is the $\alpha$ dependence of $t_T$, as exemplified
in Fig. \ref{TimeLargeSmall}. 
For regimes $0.1$ through $2.2$, the time spent in a given regime is a decreasing function of $\alpha$. The duration of regime $3.2$ is an increasing function of $\alpha$.  If follows that for any given $R$ there will be an $\alpha$ that minimizes $t_T$.
 For $R=3 \rm{AU}$ this value is  $\alpha\sim 3 \times 10^{-2}$ while for $R=8 \rm{AU}$ it is $\alpha \sim 10^{-3}$.  Provided $\alpha$ and $R$ are within the range described in Sec. 7.1, this allows us to determine the
 minimum and maximum values of $\alpha$ for which the total
growth through all regimes is less than the observationally constrained
time scale ($<10^6$ yr) of planetesimal formation.

Figure \ref{TimeVarious} is analagous
to Fig. 
\ref{TimeLargeSmall} 
but for different dust grain material densities and 
dust/gas surface densities
in the disc (see Fig. caption).

Except for regime $0.1$, equation \ref{tauE} and the Epstein regime time scale equations (\ref{t11E}, \ref{t21E} and similar) depend neither on $\rho_d$ nor $l_{mfp}$.  As our regime boundary and maximum velocity equations are cast in terms of $\tau$, the use of $\phi$ therefore makes clear the Epstein regime's lack of dependance on $\rho_d$ and $l_{mfp}$.
 However, the mass and radius of the grains  vary as $\rho_d^{-2} \phi^3$ or $\rho_d^{-1}$ 
at various stages of the evolution so the actual grain radii at which the regime transitions occur do vary.  In the Stokes 
regime, density plays a non-trivial role in the effects of turbulence on the grain trajectory (\ref{tauS}) and therefore 
in determining the boundary between the two drag regimes as well as the details of the regimes of $\frac{v_c}{H_d}$.  In figure \ref{TimeVarious} we plot the time scales for $\rho_d=0.5 \rm{gcm}^{-3}$.

From (\ref{main}), we see that  increasing the dust disc surface density $\Sigma_d$ will decrease the growth time scale.  However, in our equations $\Sigma_d$ frequently appears in a ratio with $\Sigma_g$.  The effect of changing $\Sigma_d$ while keeping $\Sigma_g$ constant is greater than that of changing both equally, as can be seen in Fig. (\ref{TimeVarious}).  The total surface density primarily affects the condition for gravitational instability which  determines the time spent in regime $3.3$. This  dominates the growth time scale at large $R$.  Changing the dust surface density alone has a strong influence throughout the disc.

\subsection{Dead zones reduce the total growth time}

The  MRI \citep{Balbus91} instability may be the primary source of turbulence in a protoplanetary 
disc but 
the entire gas disc need not be sufficiently ionized for the instability to occur.  The MRI-stable region of the disc is labeled a
dead zone (\citet{Matsumura05}, \citet{Sano00}) and will have a smaller value for $\alpha$
than the live zone.  If there is a dead zone for approximately the gas
scale height over some range of $R$, then we can simply treat the 
live and dead zones as having seperate values of $\alpha$ and
merging them in our radially depedendent calculations, 
while noting that dust unable to decouple from the turbulence on a 
viscous time scale could be deposited at the outer edge of the dead zone.

If instead we consider a turbulent disc with a dead zone near the midplane of the disc (with a height $H_D$) for some range of $R$ (but live for a significant thickness above the midplane) we can approximate the growth time scales by calculating the time scales for the live zone turbulence until the dust disc is contained within the dead zone $H_d=H_D$ at which point we proceed with the dead zone turbulence.  Noting that the time scales of regimes $2.2$ and $3.2$ depend on $\alpha$ only for their beginning and end conditions, respectively, and that the onset of regime $2.2$ is precisely the same as that of settling it follows that the dead zone $\alpha$ only effects the time scale of regime $3.2$ and so the effect is simply calculated.  Note that halfway settling the dust, $H_d=0.5 H_g$ occurs for roughly 
\begin{equation}
\tau= \frac{2\sqrt{\alpha}}{\Omega}.
\label{taud12}
\end{equation}
within regime $2.2$ (Table \ref{RegimeOrder}).
 The time scale behaviour with a dead zone is important because the duration of regime $3.2$ is an increasing function of $\alpha$, whereas regimes $0.1$ through $2.2$ are decreasing functions of $\alpha$.  Regime $2.2$'s time scale, as discussed above, will only depend on the live zone $\alpha$.  
Accordingly, as the contrast between the values of $\alpha$ in
the live and dead zones increase,  the 
the total time for planetesimal growth decreases.
In particular therefore, having a dead zone, as opposed to no dead zone,
decreases the total time below that of a disc with only one of the two values of $\alpha$.

The role of a dead zone in lowering the growth time scale
is shown  in Fig. (\ref{TimeLiveDead}).

\section{Constraints on the range of $\alpha$, $p$ and $R$ for which gravitational instability occurs}

Armed with equations for growth rates and collisional velocities 
of the dust grains, we can identify the range of values of $\alpha$ and $p$ that will 
permit planetesimal growth in a realistic 
turbulent disc on observable time scales for ranges of $R$.

While we are restricted by our regime definitions and regime ordering 
to values of $\alpha\ge 10^{-6}$, the effect of $\alpha$ on the estimated
time scales for planetesimal growth is pronounced.
We now discuss the effect of $\alpha$ and $p$ on the range of values that allow for 
planetesimal formation on the observationally relevant time scales.

\subsection{Constraints implied by the maximum  $v_c$}

We can place a maximum value for $\alpha$ that still allows planetesimal 
formation by studying its influence on the 
the maximal collisional velocity.  
We have used only the simple sticking parameter $c$
to characterize the extent of sticking between interacting dust grains.
However, for large enough values of $v_c$, the dust grains will destroy each other on impact.  
A more complete dust grain interaction model would allow us to constrain
 $v_c$   and thus $\alpha$
because  the maximum $v_c$ varies as $\sqrt{\alpha}$ from (\ref{vcmax}). For a disc with a dead zone, the maximum collisional velocity in the dead zone will be
lower than that in the live zone because of the lower value of $\alpha$. 
This helps facilitate planet formation even if the live zone $\alpha$ is otherwise too
large.  We quantify this below.

If the dust scale height for $\tau=\Omega^{-1}$ approximately defines the onset of the dead zone ($H_d = \sqrt{\alpha} H_g << H_g$ for either $\alpha$) 
 then we can find the maximum live zone collisional velocity from (\ref{vcmax}):
\begin{equation}
v_{c, max}=\sqrt{\alpha} c_s \left(1-e^{-1}\right)^2 \simeq 0.4 \sqrt{\alpha} c_s.
\label{vcmax4}
\end{equation}

If instead the end of regime $2.2$ defines the beginning of the dead zone, then from Eqs \ref{vc2} and \ref{tau22} we have 
\begin{equation}
v_{c, max}=\sqrt{\frac{\alpha}{x_{max}}} B c_s \simeq 0.35 \sqrt{\alpha} c_s,
\label{vcmax5}
\end{equation}
and a dead zone height from Eqs \ref{hd2} and \ref{tau22} of
\begin{equation}
H_D=\sqrt{\alpha} x_{max} H_g \simeq 2 \sqrt{\alpha} H_g.
\label{hd22}
\end{equation}
If the height of the dead zone is roughly half that of the gas disc, then Eq.
\ref{taud12} gives the condition for the dead zone height to
equal the dust scale height.
The maximum live zone collisional velocity occurs for that value of $\tau$ and is
\begin{equation}
v_{c, max}=\sqrt{2} \alpha^{\frac 34} B c_s \simeq 0.75 \alpha^{\frac34} c_s.
\label{vcmax6}
\end{equation}

\begin{figure}
\plottwo{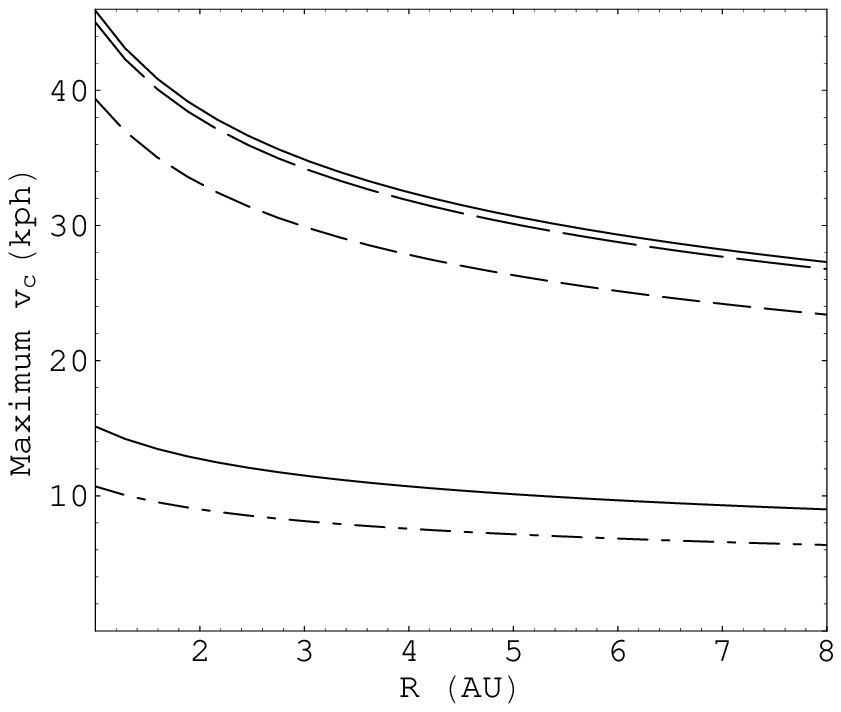}{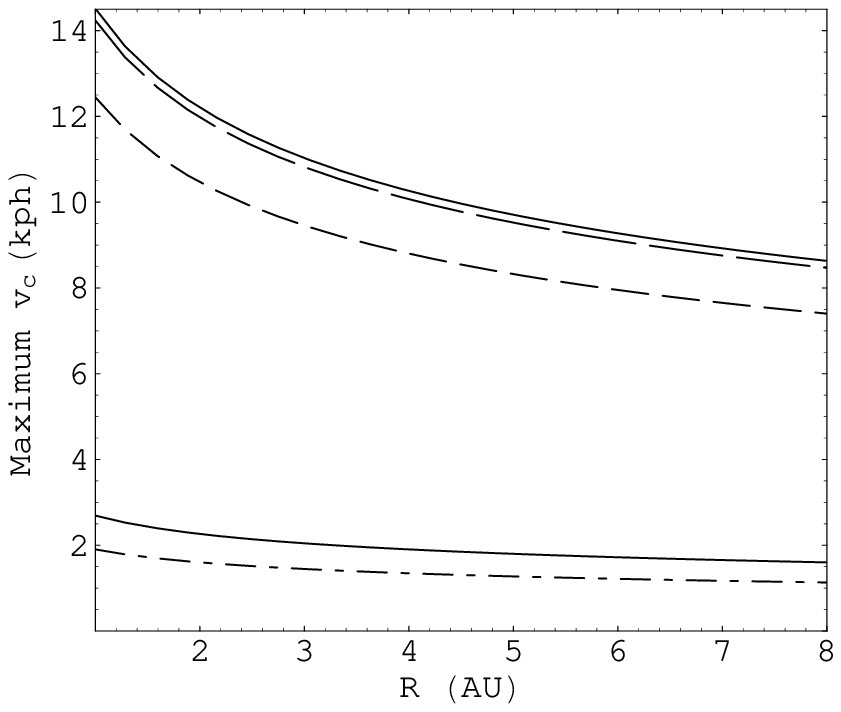}
\caption{Maximum collisional velocity in km/h as a function of distance from the star for a live zone $\alpha$ of $10^{-3}$ (left) and $10^{-4}$ (right). 
 The higher solid lines are the velocities  
from (\ref{vcmax}) for the live zone while the lower 
solid lines is the live zone velocity from (\ref{vcmax6})
for a disc with a dead zone with $H_D=0.5 H_g$.  The long dashed, short dashed and dash-dotted lines are the collisional velocities at $\tau=\Omega^{-1}$ and the ends of regimes $2.2$ and $2.1$ respectively.  Because these 
curves apply for grains of cm to meter size scales, common experience 
suggests that sold grain destruction from collisions $> 30$km/h
would lead to obliteration rather than sticking. This   
renders further growth  impossible for a live zone 
$\alpha\ge 10^{-3}$ unless there is a contemporaneous dead zone.}
\label{MaxVel}
\end{figure}

We plot results  for live zone values of 
$\alpha=10^{-3}$ and $\alpha=10^{-4}$ in Fig. (\ref{MaxVel}).  The maximum collisional velocities occur for dust grains of 
centimeter to meter scales. The top solid line in each panel
represents the maximum collisional velocities without a dead zone
and the bottom solid line represents the maximal live zone collisional
velocities for a dead zone with $H_D=0.5 H_g$.  From common experience, we infer that it is unlikely for dust grains colliding
with speeds greater than $30$km/h to stick from the collision (and more likely that they  obliterate).  This constraint is consistent with \cite{YoudinMaxVel} who finds
limits for $v_c$ near $5 \rm {ms}^{-1} \simeq 18 \rm{km/h}$
from the physics of dust grain collisions.
Therefore we interpret 
Fig. (\ref{MaxVel}) to imply that, while for a live zone 
$\alpha$ near or above $10^{-3}$,  a significant dead zone (with a lower 
$\alpha \sim 10^{-4}$) is needed in order
to avoid the collisional destruction and enable
planet formation in the absence of additional processes like vortices. 
However, if the live zone $\alpha$ were as low as $10^{-4}$ then
a dead zone would be  unecessary.

\subsection{Constraints implied by the total growth time scale through regime 3.2}

The discovery of a planet younger than a million years old \citep{d'Alessio05}
requires, at minimum, that planetesimals grow to the 
gravitational unstable phase on a time scale substantially shorter.  For $R \la 3 \rm{AU}$ and $p=1$ our total growth time scale to reach this phase, $t_T$,
satisfies $t_T\le 10^4$yr
for $10^{-3}<\alpha <10^{-4}$, allowing 
at least an order of magnitude leeway 
 while still staying well below $10^6 \rm{yr}$ (Fig. \ref{TimeLargeSmall}).  We also see that as $\alpha$ falls below $10^{-4}$ the total time scale rapidly rises.  This allows us to place a lower bound on $\alpha$ greater than the bound required by the validity (section 4.2) 
of our regime ordering in Table (\ref{RegimeOrder}).

\section{On the validity of key approximations}

\subsection{Range of $\alpha$ for which Table \ref{RegimeOrder} is applicable.}

Both the existence and order of the regimes for $v_c$ and $H_d$ (Table \ref{RegimeOrder}) can vary based on the strength of the turbulence ($\alpha$).  For example, if $\alpha=0$ then the only form of $v_c$ of ours that applies is thermal (Eq. \ref{vc0}), and our derivation of $H_d$ does not apply.

For $0.5AU<R<8AU$ no changes from the canonical
regime ordering  occur for $2\times 10^{-6}<\alpha<1/3$.  As $\alpha$ drops below about $10^{-6}$ our present regime ordering of Table 2 
breaks down as shown below, though a new ordering (and possibly new regimes) could be found through a similar approach, as long as the settling time remains short (section 3.2).

If $\alpha<x_{max}^{-1}$, then $\frac{\alpha}{\Omega}<\frac{1}{\Omega x_{max}}$.  It follows that the transition from $v_{h, 1}$ to $v_{h, 2}$ will occur before that from $v_{c, 2}$ to $v_{c, 3}$.  For large grains, a possible deviation from the bottom-to-top canonical ordering shown in Table \ref{RegimeOrder} arises if gravitational instability is triggered before $v_{c, 3}$, which for $0.5AU<R<8 AU$ requires $\alpha \la 10^{-8}$.  For small grains, possible
 deviations from the canonical regime ordering occurs when $\frac{t_m}{x_{max}}=\frac{\sqrt{\alpha}}{\Omega}$ (Sec. 4.3) and when equations \ref{phi01E}, \ref{phi01S} (see appendix) hold for $\tau>{t_m}{x_{max}}$. In this  case, the weakness of the turbulence delays the transition from thermal collisions to turbulent collisions.

\subsection{Constraints from comparing headwind induced infall and grain growth 
time scales}

We have made the approximation that $\Sigma_d(R)$ and $\Sigma_g(R)$ are 
time independent  and that dust grains do not migrate.  However, while 
gas depleted by accretion is replenished from the outer regions of the disc or envelope, dust migration cannot be so simply ignored.  The gas in a protoplanetary disc is partially pressure supported and so rotates at a sub-keplerian velocity.  The dust grains will then feel a headwind and begin to spiral in towards the star (\citet{WeidenHeadWind}, \citet{Nakagawa86}).  As seen in \citet{Nakagawa86}, this sub-keplerian rotation behaves as their eq 1.9 (which omits a fraction sign):
\begin{equation}
v_{orb, g}=(1-\eta)v_K,
\label{orbitgas}
\end{equation}
where $v_K$ is the Keplerian velocity and
\begin{equation}
\eta =-\frac{R^2}{2GM\rho_g}\frac{dP}{dR} = \frac {13}{8} \left(\frac{c_s}{\Omega R} \right)^2
\label{eta}
\end{equation}
parameterises the difference between the gas orbit speed and the Keplerian speed.  The gas disc also accretes viscously, with time scale $t_{\nu}=\frac{R^2}{\alpha c_s H_g}$ and accretion velocity $v_{\nu} \simeq \frac{\alpha c_s^2}{R \Omega}$.  If we assume $\rho_g >> \rho$, our $\tau^{-1}$ corresponds to the $D$
of \citet{Nakagawa86}, so adding in the viscous radial velocity we rewrite their eq 2.11 for the infall speed as
\begin{equation}
V_R=-2 \eta \Omega^2 R \frac{\tau}{1+\left(\Omega \tau \right)^2} - \frac{\alpha c_s^2}{R \Omega} \frac{1}{1+\left(\Omega \tau \right)^2}=v_r+v_{\nu}, 
\label{vr}
\end{equation}
where $v_r=2 \eta \Omega^2 R \frac{\tau}{1+\left(\Omega \tau \right)^2}$ is due to the headwind and $v_{\nu}= \frac{\alpha c_s^2}{R \Omega} \frac{1}{1+\left(\Omega \tau \right)^2}$ is due to the viscous accretion flow.  As the infall due to viscous accretion flow $v_{\nu} \la 1.2 \times 10^{-5}  \left({\alpha\over 10^{-3}}\right)
 \frac{\rm{AU}}{\rm{yr}}$ is too small to allow for significant infall within $10^5$ years we neglect the term.  The maximum $V_R(\tau)$ then is
$\eta \Omega R \simeq \frac {1 \rm{AU}}{89 \rm{yr}}$  
and occurs, for any $R$, at $\tau=\Omega^{-1}$.
  For small $\tau$, $V_R$ scales as $\tau$ while for large $\tau$,  $V_R$ scales as $ \tau^{-1}$.

\begin{figure}
\plotone{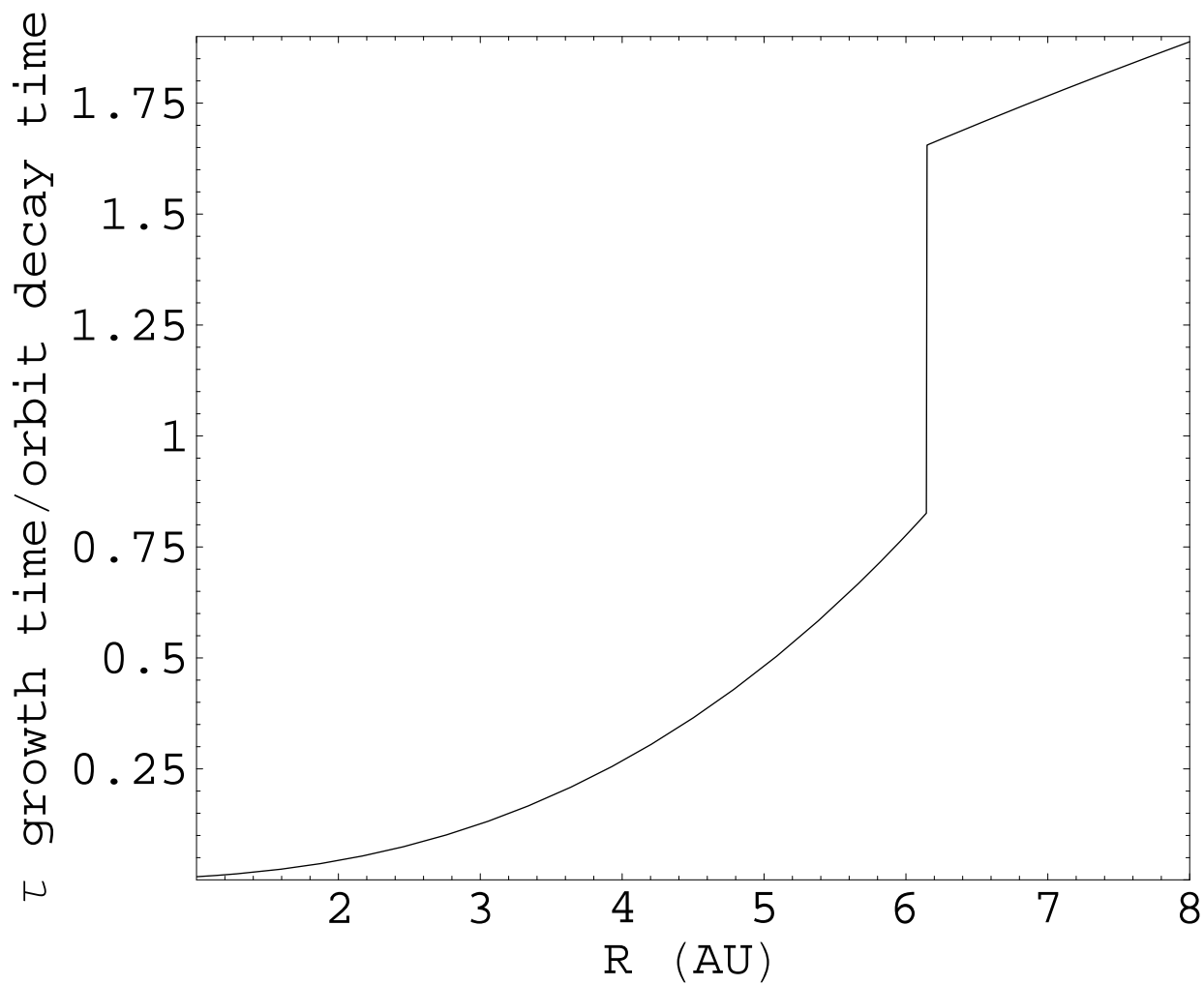}
\caption{The ratio 
 of the growth time scale $\frac{\tau}{\dot{\tau}}$ to the infall time scale $\frac{R}{v_R}$ for the headwind induced infall as calculated in section 6.1.2, 
as a function of distance from the star.  In this plot we take $p=1$,
but multiplying the curve by $1/p$ characterizes the result for other values
 of the sticking parameter.  The discontinuity at $R \simeq 6 \rm{AU}$ is due to the change in drag regime since $\tau_S \sim \phi^2$ whereas $\tau_E \sim \phi$.}
\label{HWratio}
\end{figure}

\begin{figure}
\plotone{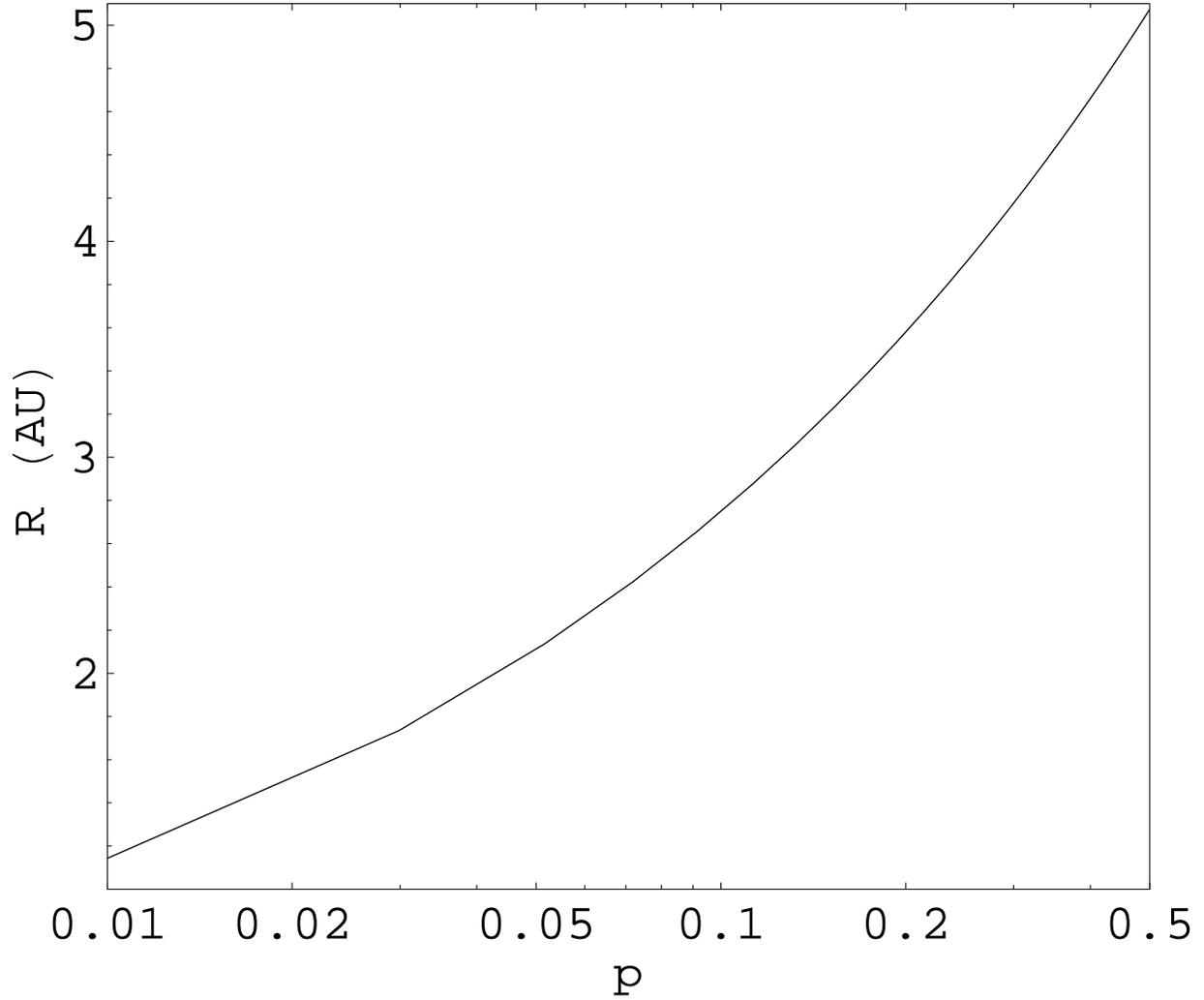}
\caption{The distance from the star in $\rm{AU}$ at which the headwind
induced infall time of Sec. 6.1.2 equals the growth time scale of $\tau$, as a function of the sticking parameter $p$.  This distance is  
an approximate measure of the radius  inside of  which the headwind 
effect can be neglected.}
\label{HWrc}
\end{figure}

Migration can be ignored only if the decceleration time scale of the infall speed $-V_R / \dot{V}_R$
is less than the infall time scale $R / V_R$ for the most affected grains.  
For $\tau>\Omega^{-1}$, $V_R \propto \tau$ and the decceleration time scale can be written as $\tau / \dot{\tau}$ for any drag regime.  We can
then cast the condition to ignore migration 
as the requirement that the ratio ${\tau / \dot{\tau} \over R/V_R}<1$. In Fig. \ref{HWratio} we plot $1/c$ times this ratio.  The jump in the plot at $R \simeq 6 \rm{AU}$ is due to the fact that $\frac{\tau_S}{\dot{\tau_S}}=\frac{\phi}{2\dot{\phi}}$ but $\frac{\tau_E}{\dot{\tau_E}}=\frac{\phi}{\dot{\phi}}$ and $R=6 \rm {AU}$ is the triple point of $\tau=\Omega^{-1}$  (sec. 5.1)
 where the drag regime for the appropriately sized grains changes (as seen in Fig. \ref{SizeBase}).  In Fig. \ref{HWrc} we plot the $R$ inside of which the headwind infall can be ignored, as a function of $p$.  It follows that
infalling grains will pile up at this radius, leading to a disc dust
density enhancement and opportunity for enhanced planetesimal formation. 
This in turn, allows more flexibility
in the value range for  $p$  that still facilitates 
rapid planetesimal formation in the absence of other physics.

There is a subtlety  however.  As we can see in Fig. (\ref{SizeBase}), 
inside of $R_{tp}$ (sec. 5.1, for our disc $R=6$AU), the curve $\tau=\Omega^{-1}$ (second curve from the top in the right panel)
lies within the Stokes drag regime, while for $R>R_{tp}$  Epstein drag applies.  In the Stokes regime, the curve is an increasing function of $R$ so 
as grains are transported inward, their $\tau$ will increase above $\Omega^{-1}$ and grain growth past the curve
will always occur.  For the Epstein drag regime however, the inward transport will keep the grains below the $\tau=\Omega^{-1}$ curve; here
 if grain growth at a given radius is slower than the inward transport, the 
grain  will never transit the curve, preventing further growth and forbidding the onset of 
gravitational instability. 

To see whether this circumstance 
arises, we calculate the rate of change of $\phi_c$
(where $\phi_c=\frac{3 \Sigma_g}{2}$ is the critical size required to satisfy $\tau_E=\Omega^{-1}$) due to the radial transport and compare it to $\frac{d\phi}{dt}$ from Eq. (\ref{main}).  We find
\begin{equation}
\frac{d\phi_c}{dt}=\frac{d}{dt} \frac{3 \Sigma_g}{2} =
-\frac32 \frac{\phi_c}{R} \frac {dR}{dt} = \frac32 \phi_c \eta \Omega R = \frac{117}{36} \frac {\Sigma_g \Omega H_g^2}{R^2}
\label{dpcdt}
\end{equation}
where $\Sigma_g$ scales as $R^{-\frac32}$ from (\ref{sigr}) and
\begin{equation}
\frac{d\phi}{dt}=p \Sigma_d \Omega \left(1-e^{-1}\right)
\label{dpdtto}
\end{equation}
for grains with $\phi=\phi_c$.  For the disc model we have chosen in Sec. 2.2., we find for $\phi=\phi_c$ that
\begin{equation}
{d\phi/dt\over d\phi_c/dt} \simeq p 1.8{\left({\rm{R}}\over{\rm{AU}}\right)^{-\frac12}}\left({\Sigma_g/\Sigma_d\over 100}\right)^{-1}.
\label{epgrowth}
\end{equation}
In that portion of the disk where this quantity is less than $1$ in the Epstein drag regime, the inward transport prevents the grains from growing above $\tau=\Omega^{-1}$.  For our canonical disc parameters this occurs everywhere that $\tau=\Omega^{-1}$ occurs for Epstein drag (i.e. $R>R_{tp}$).  Because the time scale to reach $\phi_c$ is an increasing function of $R$ while the total mass contained within an annulus of width $dR$ is a decreasing function of $R$, changes in $\Sigma_d$ as the dust falls inward do not alter the basic conclusion.  Increasing $\Sigma_g$ increases the 
triple point $R_{tp}$ of (Eq. (\ref{RTriple})) and increasing the ratio $\Sigma_g/\Sigma_d$ decreases 
the radius where (\ref{epgrowth}) drops below unity.
 Thus the  critical radius above which planetesimal growth
is prevented is indeed $R_{tp}$.
The effect remains important if the basic disc parameters are altered,  
but the radius of the triple point and 
and the critical radius where (\ref{epgrowth}) drops below unity 
and the triple point radius  (\ref{RTriple}) can change.

The important upshot of the effect just described is that 
gravitational instability cannot be easily reached at radii larger than  the 
triple point (discussed in section 5.1) of the  $\tau=\Omega^{-1}$ curve
without additional physics like vortices, 
or a significantly different disc model.  
It is interesting, if not unsurprising in hindsight, 
that the drag induced grain agglomeration benefit from
  vortices \citep{Barge95} is maximized 
 for the same $\tau = \Omega^{-1}$ condition that 
gives the maximal infall speed we found from (\ref{vr}) .

The conclusions drawn from (\ref{epgrowth}) and the limits on $p$ from Fig. (\ref{HWrc})
are essentially independent of $\alpha$
because its effects on $v_c$ and $H_d$ in 
Eq. (\ref{main}) for the relevant regimes
 cancel.   This allows us to use Fig. (\ref{HWrc}) to constrain our sticking parameter, indepedent of $\alpha$ (at least for $R<8AU$).  
This makes our constrains on $\alpha$ more meaningful.

\subsection{Neglect of gravitational enhancement to cross section}
Our model ends at the point
when gravitational instabilities take over.
We have calculated 
whether gravitational enhancement to dust-dust collisions
plays any role before that time. We 
find the effect to be small (typically under $5\%$)
for all but the largest $\alpha$ even for the largest scale dust.  

\subsection{Use of linear vs. quadratic in $v$ drag forces}

Even though the present flow is turbulent,
the laminar approximation for the Stokes drag was used because
the grains were presumed to be smaller than the smallest
turbulent eddies at all times. Therefore, unlike a macroscopic
object in a turbulent flow, the grain would see an effectively
laminar flow.  

This approximation can be justified a posteriori: 
It turns out that only very 
close to the star, do the largest scale dust grains just before gravitational
instablity become comparable to the smallest scale turbulence. 
We neglect the change in drag law that this would imply
because those  eddies both
are far less massive than the grains and are 
traveling at velocities nearly equal to these 
dust grains at the onset gravitational instability and so will have negligeable effects on the grain trajectories.

\subsection{Sticking}

The most uncertain aspect of our model lies in the sticking mechanism.  A value of $p=1$, which represences 100\% efficiency
for collisions resulting in sticking, seems  unlikely.  
Considering $p$ to be a gestalt parameter combining both all the possible sticking and destruction results of collisions and examining Eq. \ref{main} we can see that our time scale $T$ depends inversely on $p$.  Comparison with sec. 7.2 we see that decreasing $c$ strongly decreases the outer value for $R$ allowed by the headwind induced infall,
 although this will only become important if $p$ nears $10^{-2}$.  If we had a well developed sticking model we could calculate 
the maximum collisional velocity before grain collisions result in destruction.  Accordingly, our ability to constrain $\alpha$, either through disallowing grain destruction through high velocity collisions or through time scale arguments is limited by the quality of the sticking model.

\subsection{Dust size spectrum}

In our model we assume that all the dust grains share a common radius $r_d$, and in section 2 we justify this assumption by appealing to an eventual Gaussian spectrum.  As we saw in section 7.2, the differential rotation of the dust and gas discs caused by partial pressure support of the latter induces size dependent velocities in the former, the radial one given by Eq. \ref{vr} and an orbital one (\citet{Nakagawa86}):
\begin{equation}
v_{\theta}=\frac{\eta \Omega R}{1+(\Omega \tau)^2}.
\label{vtheta}
\end{equation}
An adequately broad dust size spectrum would then affect the grain-grain interactions by changing the form of $v_c$.  We can evaluate the importance of this effect by calculating the width of the dust spectrum ($\Delta \tau/\tau$) required for the difference in headwind induced velocity to equal the velocity induced by the eddies which dominate dust-dust collisions in our model from Eq. \ref{v2}.  We therefore calculate the minimum $\Delta \tau/\tau$ for which a spectrum would influence our results.  The larger the $\Delta \tau/\tau$ we calculate, the broader the spectrum our model can accommodate.  For $0.5$ AU $<R<$ $8$ AU and $10^{-4}<\alpha<10^{-2}$, we find that the minimum $\Delta \tau/\tau$ for which a spectrum would affect our results is $0.12$, in regime $3.2$.  For Epstein drag, this is also $\Delta r_d/r_d$ while for Stokes drag this implies $\Delta r_d/r_d=0.06$.

\citet{draine2006} gives an observationally derived spectrum for small grains that deviates from our Gaussian.  However, for most of the planetesimal growth, the grains sizes that we model are much larger than the micron grain sizes in the Draine spectrum.  While larger scale grain-grain collisions will produce some detritus down to the micron scales, we cease to model the spectrum on these scales once the bulk of the dust mass is predominately comprised of larger grains.  There are few observational constraints on the grain spectrum on scales much larger than microns.  This latter circumstance justifies our having pursued the simple ``dust of a single size at any one time'' approach herein.  The benefit is that the basic physics can be identified at each stage of growth, the formalism is simple, and meaningful constraints in growth times can be calculated analytically.

\section{Conclusion}

We have developed a simple model to estimate the effects of disc 
turbulence on planetesimal growth 
We show that turbulence has two competing influences: (1) It increases
the collisional velocity of dust grains which 
increases their interaction for growth. (2) It increases the velocity
which determines the scale
height of the dust layer in the disc which decreases the volume
density lowering the interaction for growth. Since the growth
rate of grains depends on the ratio of these two velocities,
the evolution of protoplanetesimal formation depends on the evolution
of this ratio. Because the coupling of the gas to the dust 
evolves with grain size, the ratio is not a constant.
Our work differs from previous studies 
by explicitly incorporting the dynamical evolution of
dust grain size and the dust settling and considering discs
with and without dead zones.

We identify the range of turbulent strength, measured by $\alpha$,
 the range of  dust sticking efficiency, measured
by $c$, and the disc radius range 
for which protoplanetesimal formation progresses to the
gravitational
instability phase on observtionally constrained time 
scales $(<10^6)$ yr, \citet{d'Alessio05}). 
The allowed ranges emerge from constraints on 
(1) the total time scale for planetesimal formation, (2) the infall time scale of the dust grains compared to their growth time and (3) the destruction of dust grains from high velocity collisions. We find that 
 dust grain infall due to headwind gas drag prohibits reaching the 
gravitational instability phase  
outside a limiting radius 
that depends on the transition between Stokes and Epstein drag (the $\tau=\Omega^{-1}$ triple point from sec. 5.1).
For  our disc model this occurs at $R = R_{tp}$ given by  Eq. (\ref{RTriple}).  
It is possible that material piles up at $R=R_{tp}$ which may
enhance planetesimal growth at this radius. This latter
effect requires further work.

For $R<R_{tp}$, we find that for 
$p>10^{-2}$ and $\alpha \le 10^{-3}$ turbulence does not catastrophically   
 slow the needed planetesimal growth, 
as seen in Fig. (\ref{TimeLargeSmall}), 
even though the time and size scales at which the growth occurs still depend strongly on those parameters. These results depend on the unknown physics of grain-grain interactions
determining $p$, but the assumption that grain destruction results from 
 collisions with 
velocities greater than $30$km/hr strictly requires $\alpha < 10^{-2}$ for all discs 
if growth is to be fast enough. As determined in Sec. 6.1.1, values of 
$\alpha \simeq 10^{-3}$ for discs with substantial dead zones and $\alpha \simeq 10^{-4}$ are acceptable.

The condition that the grain infall time
be longer than the grain growth time (as discussed in section 7.2)
also leads to an $\alpha$ independent minimum acceptable $p$ as seen
in Fig. (\ref{HWrc}).  We find that as $p$ falls below $\frac{1}{10}$ the distance from the star at which gravitational instability can be reached and planetesimal formation occurs falls below $3$ AU.  For $p=10^{-2}$ planetesimal formation won't occur outside of roughly $1$ AU.  Values of $c$ below $10^{-2}$ are too low
to produce rapid enough planetesimal growth 
in the absence of additional physics.

Although more 
detailed calculations with grain size spectra, collisional velocity spectra, 
and a  more detailed grain-grain collision model are needed,
our most robust conclusion is that 
planetesimal growth in an initially gravitationally stable turbulent disc 
is not prohibited by  turbulence
with $10^{-6}\le \alpha \le 10^{-3}$, even in the absence
of including dust agglomeration enhancement mechanisms.
 For $\alpha \ge 10^{-2}$ we 
do not find much opportunity 
 for growth, even for discs with a dead zone,
and additional physics to promote dust growth such as large scale vortices,
or the consequences of mass pile up at $R=R_{tp}$ 
need to be considered.

\section{Appendix}

Here we derives the various formulas for Tables \ref{RegimeSize} and \ref{RegimeTime}.

\subsection{Regime $0.1$}

In this regime  we have $H_d=H_{d, 1}=c_s \Omega^{-1}$ from (\ref{hd1}) and $v_c=v_{c, 0}=v_{th}= \sqrt{\frac{m_g}{m_d}}c_s$ from (\ref{vc0}).  Here 
the  dust motion is sufficiently strongly coupled to the bulk motion of the gas that the turbulent eddies have only  the effect of preventing settling.  Accordingly, (\ref{main}) becomes
\begin{equation}
\frac {d\phi}{dt}=p \Sigma_d \Omega \sqrt{\frac{m_g}{m_d}}=p \Sigma_d \Omega \sqrt{\frac{3m_g}{4\pi \phi^3}} \rho_d.
\label{dpdt1}
\end{equation}
Solving (\ref{dpdt1}), the time scale $t_1(\phi_f)$ for regime $1$ to grow dust grains from $\phi_i$ to $\phi_f$ is:
\begin{equation}
t_{0.1}(\phi_f, \phi_i)=\frac {4}{5p} \sqrt{\frac{\pi}{3m_g}} \frac{1}{\Sigma_d \Omega \rho_d}\left(\phi_f^{\frac 52}-\phi_i^{\frac 52} \right) \simeq \frac 45 \sqrt{\frac{\pi}{3m_g}} \frac{1}{\Sigma_d \Omega \rho_d} \phi_f^{\frac 52} = t_{0.1}(\phi_f).
\label{t01}
\end{equation}
Because the growth rate in this regime falls off strongly with grain size,
the time spent in the regime depends only on its final size.  This allows us to neglect any uncertainty in the initial radii of interstellar dust (which we will therefore approximate  to be $0$).  We can also see that, as suggested in Sec. 2.1, the actual size dust grains can  reach in 
 regime $0.1$  is strongly limited  by time because $t \sim \phi^{\frac 52}$.

For the parameters we consider, regime $0.1$ will lead into regime $1.1$ and 
thus the former  ends when $v_{c, 0}=v_{c, 1}$, 
after which $v_c=v_{c, 1}$.  Setting $v_c$ from (\ref{vc0}) and (\ref{vc1}) equal we can see that this occurs for
\begin{equation}
\phi_f= \phi_{0.1, E}\equiv\left(\frac{27 m_g \rho_d^2 \Sigma_g^2 t_m \Omega}{16 \pi \alpha}\right)^{\frac 15}
\label{phi01E}
\end{equation}
for Epstein drag and for
\begin{equation}
\phi_f=\phi_{0.1, S}\equiv\left(\frac{27 m_g l_{mfp}^2 \Sigma_g^2 \Omega \rho_d^4 t_m}{16 \pi \alpha} \right)^{\frac 17}
\label{phi01S}
\end{equation}
for  Stokes drag.

\subsection{Regime 1.1}
When (\ref{vc1}) is the dominant value for $v_c$ and the dust is too small to settle ($H_d=H_{d, 1}$ from (\ref{hd1})) we are in regime $1.1$.  In this regime
$v_c$ is dominated by  the smallest scale turbulent eddies,
 approaching the corresponding eddy speed as the dust grows.

Solving (\ref{main}) in this regime we find the time scale $t_{1.1}(\phi_i, \phi_f)$ to grow the grains from $\phi_i$ to $\phi_f$ is
\begin{equation}
t_{1.1, E}(\phi_i, \phi_f)=\frac {3}{2p} \frac {\Sigma_g}{\Sigma_d} \sqrt{\frac {t_m}{\alpha \Omega}} {Ln}\left(\frac{\phi_f}{\phi_i}\right)
\label{t11E}
\end{equation}
for Epstein drag and
\begin{equation}
t_{1.1, S}(\phi_i, \phi_f)=\frac{3 \Sigma_g l_{mfp} \rho_d}{2 p \Sigma_d}\sqrt{\frac{t_m}{\alpha \Omega}} \left( \frac {1}{\phi_i}-\frac{1}{\phi_f} \right) \simeq \frac{3 \Sigma_g l_{mfp} \rho_d}{2 \Sigma_d}\sqrt{\frac{t_m}{\alpha \Omega}} \frac {1}{\phi_i}
\label{t11S}
\end{equation}
for Stokes drag.  Note that these values for $t_{1.1}$ are not strongly dependent on $\phi_f$, allowing us to use the approximations implicit in (\ref{vc1}).

Once the damping time grows
long enough such that 
\begin{equation}
\tau=\frac {t_m}{x_{max}}
\label{tau11}
\end{equation}
 then the "Goldilocks" turbulent scale exists, $v_c=v_{c, 2}$ (\ref{vc2}) applies and the regime
1.1 progresses to regime 2.1.  
This occurs for
\begin{equation}
\phi_f=\phi_{1.1, E}\equiv\frac {3 \Sigma_g \Omega t_m}{2 x_{max}}
\label{phi11E}
\end{equation}
with Epstein drag and for
\begin{equation}
\phi_f=\phi_{1.1, S}\equiv\left( \frac {3 l_{mfp} \Sigma_g \Omega \rho_d t_m}{2 x_{max}} \right)^{\frac 12}.
\label{phi11S}
\end{equation}
with Stokes drag.  

\subsection{Regime 2.1}

In regime $2.1$, the dust grows to
couple with progressively  larger and larger scale eddies
that determine  the  collisional velocity $v_c$.
For the range of $\alpha$ and $R$ that we consider, 
the grains still 
remain small enough to avoid settling. 
(For smaller $\alpha$ 
and larger  $R$, 
even the largest eddies  would be be too weak to prevent  settling
and an additional Regime $1.2$ would occur.)
 
 In regime $2.1$, 
we have $v_{c, 2}$ from (\ref{vc2}) and $H_{d, 1}$ from (\ref{hd1}).  Solving (\ref{main}) we find  the growth time scale  in this regime to be
\begin{equation}
t_{2.1, E}(\phi_i, \phi_f)=\sqrt{\frac{6 \Sigma_g}{\alpha}} \frac {1}{p \Sigma_d \Omega B} \left(\phi_f^{\frac 12}-\phi_i^{\frac 12} \right) \simeq \sqrt{\frac{6 \Sigma_g}{\alpha}} \frac {1}{\Sigma_d \Omega B} \phi_f^{\frac 12}
\label{t21E}
\end{equation}
for Epstein drag and
\begin{equation}
t_{2.1, S}(\phi_i, \phi_f)= \sqrt{\frac{3 l_{mfp} \Sigma_g \rho_d}{2 \alpha}} \frac {1}{p \Sigma_d \Omega B} {Ln} \left(\frac{\phi_f}{\phi_i} \right)
\label{t21S}
\end{equation}
for Stokes drag.

Regime $2.1$ progresses to regime $2.2$ when the grains grow large enough that the turbulence can no longer prevent settling at which point (\ref{hd2}) applies.  
This occurs for
\begin{equation}
\tau=\frac {\sqrt{\alpha}}{\Omega}
\label{tau21}
\end{equation}
Accordingly Regime $2.1$ ends for
\begin{equation}
\phi_f=\phi_{2.1, E}\equiv\frac {3\sqrt{\alpha}\Sigma_g}{2}
\label{phi21E}
\end{equation}
with Epstein drag and for
\begin{equation}
\phi_f=\phi_{2.1, S}\equiv\left( \frac {3 \sqrt{\alpha} l_{mfp} \Sigma_g \rho_d}{2} \right)^{\frac 12}.
\label{phi21S}
\end{equation}
with Stokes drag.  

\subsection{Regime 2.2}

Regime $2.2$ occurs for dust grains 
large enough to  begin settling out of the gas disc and also have  
their collisional velocity determined by coupling to  non-minimum scale
 turbulence.  Here we use $v_{c, 2}$ from (\ref{vc2}) and $H_{d,2}$ from (\ref{hd2}) while approximating $1-e^{-\frac{1}{\Omega \tau}} = 1$.  Solving (\ref{main}) we find the time scale to grow from some $\phi_i$ to $\phi_f$ 
in this regime to be
\begin{equation}
t_{2.2, E}(\phi_i, \phi_f) = \frac{2}{p \Sigma_d \Omega B} \left( \frac {3 \Sigma_g}{2} \right)^{\frac 32} \left( \phi_i^{-\frac 12}-\phi_f^{-\frac 12} \right) \simeq \frac{2}{\Sigma_d \Omega B} \left( \frac {3 \Sigma_g}{2} \right)^{\frac 32} \phi_i^{-\frac 12}
\label{t22E}
\end{equation}
with Epstein drag and
\begin{equation}
t_{2.2, S}=\frac {1}{2 p \Sigma_d \Omega B} \left( \frac{3 l_{mfp} \Sigma_g \rho_d}{2} \right)^{\frac 32} \left(\phi_i^{-2}-\phi_f^{-2} \right) \simeq \frac {1}{2 \Sigma_d \Omega B} \left( \frac{3 l_{mfp} \Sigma_g \rho_d}{2} \right)^{\frac 32} \phi_i^{-2}
\label{t22S}
\end{equation}

Regime $2.2$ ends when the turbulent length scale on which 
the dust's collisional velocity depends reaches the largest turbulent scale.  This happens when $x=\frac{t_M}{\tau}=x_{max}$, which occurs for
\begin{equation}
\tau=\frac{1}{\Omega x_{max}}.
\label{tau22}
\end{equation}
Accordingly, we find that in the Epstein regime, regime $2.2$ ends for
\begin{equation}
\phi_f=\phi_{2.2, E}\equiv\frac{3 \Sigma_g}{2 x_{max}}
\label{phi22E}
\end{equation}
and 
\begin{equation}
\phi_f=\phi_{2.2, S}\equiv\left( \frac{3 l_{mfp} \Sigma_g \rho_d}{2 x_{max}} \right)^{\frac 12}
\label{phi22S}
\end{equation}
in the Stokes regime.  None of (\ref{t22E}), (\ref{t22S}), (\ref{phi22E}) or (\ref{phi22S}) depend on $\alpha$ so the only $\alpha$ dependence of the time scales that result from regime $2.2$ derives from  the initial size condition that determines when
the regime applies.

\subsection{Regime 3.2}

In regime $3.2$ we use $v_{c, 3}$ from (\ref{vc3}) while continuing to use
 $H_{d, 2}$ from (\ref{hd2}): the dust is decoupling from the turbulence while continuing to settle.  Writing (\ref{main}) in this regime we find
 \begin{equation}
 \frac{d\phi}{dt}=p\Sigma_d\Omega^3\tau^2 \left(1-e^{-\frac{1}{\Omega \tau}}\right)^2.
 \label{dpdt32}
 \end{equation}
While the above equation is not simply solvable and in our plots we integrate numerically, Regime $3.2$ can be divided into two subregimes $a$ and $b$ which admit the approximate forms:
\begin{equation}
\frac{d\phi}{dt} \ga p \Sigma_d \Omega^3 \tau^2 \left(1-e^{-1}\right)^2
\label{dpdt32first}
\end{equation}
for $\tau<\Omega^{-1}$ and
\begin{equation}
\frac{d\phi}{dt} \simeq p \Sigma_d \Omega
\label{dpdt32second}
\end{equation}
for $\tau>\Omega^{-1}$.

Using the first time scale approximation we find that the approximate time scales to grow from some $\phi_i$ to $\phi_f<\phi(\tau=\Omega^{-1})$
 in this regime are less than
\begin{equation}
t_{3.2a, E} = \frac{9 \Sigma_g^2}{4 p \Sigma_d \Omega \left(1-e^{-1}\right)} \left(\phi_i^{-1}-\phi_f^{-1}\right)
\label{t32aE}
\end{equation}
for Epstein drag and
\begin{equation}
t_{3.2a, S}=\frac{3 \Sigma_g^2 l_{mfp}^2 \rho_d^2}{4 p \Sigma_d \Omega \left(1-e^{-1}\right)}\left(\phi_i^{-3}-\phi_f^{-3}\right) \simeq \frac{3 \Sigma_g^2 l_{mfp}^2 \rho_d^2}{4 \Sigma_d \Omega \left(1-e^{-1}\right)}\phi_i^{-3} 
\label{t32aS}
\end{equation}
for Stokes drag.  Using the second time scale approximation, we find that the approximate time to grow from some $\phi_i>\phi(\tau=\Omega^{-1})$ to $\phi_f$ is
\begin{equation}
t_{3.2b} = \frac{\phi_f-\phi_i}{p \Sigma_d \Omega} \simeq \frac {\phi_f}{\Sigma_d \Omega} .
\label{t32b}
\end{equation}

Regime $3.2$ ends when the dust disc becomes gravitationally unstable, which 
occurs when the Toomre criterion \citep{Toomre}
\begin{equation}
H_d \le \frac {\pi}{2} \frac{G \Sigma_d}{\Omega^2}
\label{R33End}
\end{equation}
is satisfied.  This condition can be rewritten as
\begin{equation}
\tau= \frac {2 \sqrt{\alpha} c_s}{\pi G \Sigma_d}.
\label{tauG}
\end{equation}
Regime $3.2$ then ends when 
\begin{equation}
\phi_f = \phi_{3.2, E}\equiv\frac {3 \sqrt{\alpha} c_s \Omega}{\pi G} \frac {\Sigma_g}{\Sigma_d}
\label{phi32E}
\end{equation}
for Epstein drag
and when
\begin{equation}
\phi_f = \phi_{3.2, S}\equiv \left( \frac{3 \sqrt{\alpha} l_{mfp} \Omega c_s}{\pi G} \frac {\Sigma_g}{\Sigma_d} \rho_d \right)^{\frac 12}
\label{phi32S}
\end{equation}
for Stokes drag.

{\bf Acknowledgments}:
We acknowledge support from NSF grants AST-0406799 and AST-0406823,
NASA
grant ATP04-0000-0016, and the KITP of UCSB, where this research was
supported in part by NSF Grant PHY-9907949. 
AH acknwoledges a Horton Graduate Fellowship from the Laboratory for
Laser Energetics.

\clearpage
\begin{deluxetable}{lcc}
\tablewidth{0pt}
\tablecaption{Key Variables}
\tablehead{
\colhead{\rm{Variable}}        &
\colhead{\rm{Description}} 
}
\startdata

$\rho$ & Dust disc density \\
$\rho_d$ & Dust grain density \\
$\rho_g$ & Gas disc density \\
$\Sigma_d$ & Dust disc surface density \\
$\Sigma_g$ & Gas disc surface density \\
$\Sigma_{g0}$ & Gas disc surface density at $R=1$AU \\
$r_d$ & Grain radius \\
$\phi$ & Grain size parameter \\
$m_d$ & Grain mass \\
$m_g$ & Mass of a gas molecule \\
$v_{th}$ & Dust thermal speed \\
$\tau_E$ & Grain frictional stopping time for Epstein drag \\
$\tau_S$ & Grain frictional stopping time for Stokes drag \\
$\Omega$ & Keplerian angular velocity \\
$H_g$ & Gas disc scale height \\
$H_d$ & Dust disc scale height \\
$H_D$ & Dead zone height \\
$c_s$ & Gas sound speed \\
$p$ & Sticking parameter \\
$\alpha$ & Shakura-Sunyaev viscosity parameter \\
$\sigma_{gg}$ & Neutral gas collisional cross-section \\
$\sigma_{gd}$ & Dust-gas collisional cross-section \\
$\sigma_{dd}$ & Grain-grain collisional cross-section \\
$R$ & Distance from the star \\
$t_{T}$ & Total time scale to achieve gravitation instability

\enddata 
\end{deluxetable}

\clearpage
\begin{deluxetable}{lccccc}
\tablewidth{0pt}
\tablecaption{Regime Descriptions and Ordering \label{RegimeOrder}}
\tablehead{
\colhead{\rm{Regime}}        &
\colhead{\rm{Start Condition}}   &
\colhead{\rm{End Condition}}   &
\colhead{\rm{Collisional Velocity}}     &
\colhead{\rm{Dust Height}}
}
\startdata

Regime $3.2$ & $\tau=\frac{1}{\Omega x_{max}}$ & $\tau=\frac{2 \sqrt{\alpha} c_s}{\pi G \Sigma_d}$ & $\sqrt{\alpha} c_s \Omega \tau \left(1-e^{-\frac{1}{\Omega \tau}}\right)^2 $& $ \frac {\sqrt{\alpha} c_s}{\Omega^2 \tau} $\\ \\
Regime $2.2$ & $\tau=\frac{\sqrt{\alpha}}{\Omega} $& $\tau=\frac{1}{\Omega x_{max}}$ &$\sqrt{\alpha \Omega \tau} B c_s$ &$\frac{\sqrt{\alpha} c_s}{\Omega^2 \tau} $\\ \\
Regime $2.1$ & $\tau=\frac{t_m}{x_{max}}$ & $\tau=\frac{\sqrt{\alpha}}{\Omega}$ & $\sqrt{\alpha \Omega \tau} B c_s$ & $\frac {c_s}{\Omega}$ \\ \\
Regime $1.1$ & Eq. (\ref{phi01E}), (\ref{phi01S}) & $\tau=\frac{t_m}{x_{max}} $& $\sqrt{\frac{\alpha \Omega}{t_m}} \tau c_s$ & $\frac {c_s}{\Omega}$ \\ \\
Regime $0.1$ & Interstellar Dust & Eq. (\ref{phi01E}), (\ref{phi01S}) &$\sqrt{\frac{3 \rho_d^2 m_g}{4 \pi \phi^3}} c_s $& $\frac{c_s}{\Omega} $
\enddata
\tablecomments{The planetesimal growth regimes that occur in our model which progress
from bottom to top as the dust grains grow.
.  The collisional velocities are developed in equations (\ref{vc0}), (\ref{vc1}), (\ref{vc2}) and (\ref{vc3}) and the scale heights in (\ref{hd1}) and (\ref{hd2}).  We use these values to calculated growth rates by plugging the collisional and height velocities into eq. (\ref{main}).  The beginning and end conditions define the limits of integration for the time scale calculations.}  
\end{deluxetable}

\clearpage
\begin{deluxetable}{lccc}
\tablewidth{0pt}
\tablecaption{Regime Dust Sizes \label{RegimeSize}}
\tablehead{
  \colhead{\rm{Regime}} &
  \colhead{\rm{Final $\phi_E$}} &
  \colhead{\rm{Final $\phi_S$}} 
}
\startdata

Regime $3.2$b & $\frac {3 \sqrt{\alpha} c_s \Omega}{\pi G} \frac {\Sigma_g}{\Sigma_d}$ & $\left( \frac{3 \sqrt{\alpha} l_{mfp} \Omega c_s}{\pi G} \frac {\Sigma_g}{\Sigma_d} \rho_d \right)^{\frac 12}$ \\
\\
Regime $3.2$a & $\frac {3 \Sigma_g}{2}$ & $\left(\frac{3 l_{mfp} \Sigma_g \rho_d}{2}\right)^{\frac 12}$ \\
\\
Regime $2.2$ & $\frac{3 \Sigma_g}{2 x_{max}}$ & $\left( \frac{3 l_{mfp} \Sigma_g \rho_d}{2 x_{max}} \right)^{\frac 12}$ \\
\\
Regime $2.1$ & $\frac {3\sqrt{\alpha}\Sigma_g}{2}$ & $\left( \frac {3 \sqrt{\alpha} l_{mfp} \Sigma_g \rho_d}{2} \right)^{\frac 12}$ \\
\\
Regime $1.1$ & $\frac {3 \Sigma_g \Omega t_m}{2 x_{max}}$ & $\left( \frac {3 l_{mfp} \Sigma_g \Omega \rho_d t_m}{2 x_{max}} \right)^{\frac 12}$ \\
\\
Regime $0.1$ & $\left(\frac{27 m_g \rho_d^2 \Sigma_g^2 t_m \Omega}{16 \pi \alpha}\right)^{\frac 15}$ & $\left(\frac{27 m_g l_{mfp}^2 \Sigma_g^2 \Omega \rho_d^4 t_m}{16 \pi \alpha} \right)^{\frac 17}$
\enddata
\tablecomments{Values of $\phi$ bounding the various regimes for both Epstein and Stokes drag.  Derives from the bounding values for $\tau$ in Table \ref{RegimeOrder}.}
\end{deluxetable}

\clearpage
\begin{deluxetable}{lccc}
\tablewidth{0pt}
\tablecaption{Regime Time Scales \label{RegimeTime}}
\tablehead{
  \colhead{\rm{Regime}} &
  \colhead{\rm{Epstein Time}}     &
  \colhead{\rm{Stokes Time}}
}
\startdata

Regime $3.2$b & $\frac{\phi_f-\phi_i}{\Sigma_d \Omega}$ & As Epstein \\
\\
Regime $3.2$a & $\frac{9 \Sigma_g^2}{4 p \Sigma_d \Omega \left(1-e^{-1}\right)} \left(\phi_i^{-1}-\phi_f^{-1}\right)$ & $\frac{3 \Sigma_g^2 l_{mfp}^2 \rho_d^2}{4 \Sigma_d \Omega \left(1-e^{-1}\right)}\left(\phi_i^{-3}-\phi_f^{-3}\right)$ \\
\\
Regime $2.2$ & $\frac{2}{\Sigma_d \Omega B} \left( \frac {3 \Sigma_g}{2} \right)^{\frac 32} \left( \phi_i^{-\frac 12}-\phi_f^{-\frac 12} \right)$ & $\frac {1}{2 \Sigma_d \Omega B} \left( \frac{3 l_{mfp} \Sigma_g \rho_d}{2} \right)^{\frac 32} \left(\phi_i^{-2}-\phi_f^{-2} \right)$ \\
\\
Regime $2.1$ & $\sqrt{\frac{6 \Sigma_g}{\alpha}} \frac {1}{\Sigma_d \Omega B} \left(\phi_f^{\frac 12}-\phi_i^{\frac 12} \right)$ & $\sqrt{\frac{3 l_{mfp} \Sigma_g \rho_d}{2 \alpha}} \frac {1}{\Sigma_d \Omega B} {Ln} \left(\frac{\phi_f}{\phi_i} \right)$  \\
\\
Regime $1.1$ & $\frac {3}{2} \frac {\Sigma_g}{\Sigma_d} \sqrt{\frac {t_m}{\alpha \Omega}} {Ln}\left(\frac{\phi_f}{\phi_i}\right)$ & $\frac{3 \Sigma_g l_{mfp} \rho_d}{2 \Sigma_d}\sqrt{\frac{t_m}{\alpha \Omega}} \left( \frac {1}{\phi_i}-\frac{1}{\phi_f} \right)$ \\
\\
Regime $0.1$ & $\frac 45 \sqrt{\frac{\pi}{3m_g}} \frac{1}{\Sigma_d \Omega \rho_d} \phi_f^{\frac 52}$ & As Epstein
\enddata
\tablecomments{Equations for the time for growth from $\phi=\phi_i$ to $\phi=\phi_f$ in the various regimes, presuming $p=1$ (optimal sticking), derived by integrating Eq. \ref{main}.  The time scales are inversely dependent on $p$ and are derived in the appendix.}
\end{deluxetable}

\clearpage

\clearpage

\clearpage

\clearpage

\clearpage

\clearpage

\clearpage

\clearpage

\end{document}